\documentclass[fleqn,10pt]{wlscirep}
\usepackage[utf8]{inputenc}
\usepackage[T1]{fontenc}
\usepackage{bm}

\usepackage{amsmath}
\usepackage{graphicx}

\graphicspath{ {./imgs/} } 

\newcommand{\fTVB}{TVB C++} 

\usepackage{caption}
\usepackage{subcaption}
\usepackage{listings}

\definecolor{codegreen}{rgb}{0,0.6,0}
\definecolor{codegray}{rgb}{0.5,0.5,0.5}
\definecolor{codepurple}{rgb}{0.58,0,0.82}
\definecolor{backcolour}{rgb}{0.95,0.95,0.92}

\usepackage{color}
\usepackage[normalem]{ulem}		

\lstdefinestyle{mystyle}{
  backgroundcolor=\color{backcolour}, commentstyle=\color{codegreen},
  keywordstyle=\color{magenta},
  numberstyle=\tiny\color{codegray},
  stringstyle=\color{codepurple},
  basicstyle=\ttfamily\footnotesize,
  breakatwhitespace=false,         
  breaklines=true,                 
  captionpos=b,                    
  keepspaces=true,                 
  numbers=left,                    
  numbersep=5pt,                  
  showspaces=false,                
  showstringspaces=false,
  showtabs=false,                  
  tabsize=2
}

\title{\fTVB{}: A Fast and Flexible Back-End for The Virtual Brain}

\author[1]{Ignacio Martín}
\author[2]{Gorka Zamora-López}
\author[3]{Jan Fousek}
\author[4]{Michael Schirner}
\author[4]{Petra Ritter} 
\author[3]{Viktor Jirsa}
\author[2,5]{Gustavo Deco}
\author[1,2,*]{Gustavo Patow}

\affil[1]{\textit{ViRVIG, Universitat de Girona, Girona, Spain}} 
\affil[2]{\textit{Center for Brain and Cognition, Computational Neuroscience Group, Department of Information and Communication Technologies, Universitat Pompeu Fabra, Spain}} 
\affil[3]{\textit{Institut de Neurosciences des Systèmes, Aix Marseille Université, Marseille, France}} 
\affil[4]{\textit{Berlin Institute of Health, Charité – Universitätsmedizin Berlin, Germany}} 
\affil[5]{\textit{Institució Catalana de la Recerca i Estudis Avançats (ICREA), Barcelona, Spain}} 
\affil[*]{e-mail: gustavo.patow@udg.edu}

\begin{abstract}
This paper introduces \fTVB{}, a streamlined and \emph{fast} \emph{C++} Back-End for The Virtual Brain (TVB), a renowned platform and a benchmark tool for full-brain simulation. \fTVB{} is engineered with speed as a primary focus while retaining the flexibility and ease of use characteristic of the original TVB platform. Positioned as a complementary tool, TVB serves as a prototyping platform, whereas \fTVB{} becomes indispensable when performance is paramount, particularly for large-scale simulations and leveraging advanced computation facilities like supercomputers. Developed as a TVB-compatible Back-End, \fTVB{} seamlessly integrates with the original TVB implementation, facilitating effortless usage. Users can easily configure \fTVB{} to execute the \emph{same} code as in TVB but with enhanced performance and parallelism capabilities.
\end{abstract}
\keywords{The Virtual Brain, Whole-Brain model, C++, Fast}

\begin{document}

\flushbottom
\maketitle
\thispagestyle{empty}

\section{Introduction}
In recent years, technological advancements in noninvasive neuroimaging have opened unprecedented windows into the inner workings of the human brain. Techniques such as EEG, MEG, fMRI, and PET have revolutionized our understanding of perception, cognition, and behavior, enabling detailed analyses from the microscopic to the macroscopic scale.

Within this landscape, computational tools have played a pivotal role in unraveling the complexities of neural dynamics. While various frameworks exist, including those focusing on microscopic neuron models~\cite{Gewaltig:NEST,Stimberg2019:BRIAN,DuraBernal2019:NetPyNE,Hines2001:NEURON,Migliore2006:NEURON} or mesoscopic systems~\cite{Bekolay2014:NENGO,Dai2020:BrainModelingToolKit}, the emphasis on whole-brain modeling has led to the emergence of platforms like The Virtual Brain (TVB)~\cite{SanzLeon2013,Ritter2013,SanzLeon2015,Schirner2022,vanderVlag2022,vanderVlag2023}. TVB has garnered widespread recognition for its ability to simulate, analyze, and infer neurophysiological mechanisms across different brain scales.

Despite its great versatility, TVB faces a competitive challenge in terms of computational speed when compared to other simulation platforms, especially when dealing with extensive networks or large-scale simulations. 
Several approaches have been proposed in recent years to boost the performance of TVB, e.g., by porting specific models into the C-language [CITE] or interfacing with CUDA [CITE]. However, these attempts lacked the flexibility of TVB because they were restricted to specific scientific questions they wanted to address, without resulting in a general setting with the flexibility TVB has.

Here, we introduce \fTVB{}, a groundbreaking addition to the TVB family aimed at addressing these limitations simultaneously. Developed upon the powerful C++ programming language and the advanced \emph{Eigen} numeric library \cite{eigenweb}, \fTVB{} offers great computational performance without compromising on flexibility or ease of use. Unlike previous efforts ---focused solely on performance---\fTVB{} takes a holistic approach by retaining the same structure of the original TVB platform. This allows for \fTVB{} to seamlessly integrate with TVB through Python bindings and, therefore, effortlessly complement existing workflows, offering a swift and efficient backend solution for bulk simulations where computational time is critical.
With \fTVB{}, researchers gain access to a cutting-edge tool that empowers them to explore the intricacies of large-scale brain dynamics or conduct extensive parameter explorations, \fTVB{} stands ready to help advance the field, offering unprecedented speed, performance, and versatility.

The paper is organized as follows. Section~\ref{sec:background} provides a brief overview of existing computational tools to simulate neural and brain activity at different scales. In Section~\ref{sec:TVBPython} we present the original TVB implementation, its purpose, and general structure, and in Section~\ref{sec:TVB C++} we present \fTVB{} in detail, with all its input data formats, code structures, models, observables, and, finally, its Python bindings. In Section~\ref{sec:useCases} we present some use cases for \fTVB{}, along with the respective results obtained, and finally we summarize the impact and possibilities of \fTVB{} in Section~\ref{sec:Conclusions}.

\section{Background}    \label{sec:background}

In recent decades, advances in neuroimaging have facilitated unprecedented levels of analysis across various scales, ranging from the microscopic realm, which entails simulating the complex interplay among ionic channels, neurotransmitters, receptors, and action potentials, to the macroscopic scale encompassing whole-brain simulations.

Comprehensive investigations enable the mechanistic comprehension of population dynamics, contributing to the elucidation of complex structures such as Resting State Networks~\cite{ThomasYeo2011,Allen2012,Hansen2015525}, cognition~\cite{Kringelbach2015,Deco2023}, and pathologies like Epilepsy~\cite{Jirsa2017,Makhalova2022}, Stroke~\cite{Adhikari2017,Rocha2022,Idesis2023}, or Alzheimer's Disease~\cite{Stefanovski2019,Patow2023}. It is within the realm of whole-brain simulations that true integration occurs, leading to the emergence of cognition. Consequently, these models offer a panoramic view of neuronal dynamics at the global macroscopic level. Computational tools have evolved to simulate all these effects, both at the algorithmic and at the computational power levels. This has enabled the simulation of large neuronal models that, in turn, enabled accurate predictions illuminating both theoretical and clinical aspects of our brains. Among the different software frameworks developed to this end, we can distinguish three big groups: 
\begin{itemize}
    \item Class I, or microscopic-level simulators, which aim at simulating the activity of individual neurons with high detail, such as NEST~\cite{Gewaltig:NEST}, Brian~\cite{Stimberg2019:BRIAN}, NetPyNE~\cite{DuraBernal2019:NetPyNE}, and NEURON~\cite{Hines2001:NEURON,Migliore2006:NEURON}. These frameworks have proven particularly useful for simulating large networks of spiking neurons. In particular, NEST and Brian had their main target on point neurons, while NetPyNE and NEURON focused primarily on morphologically extended neurons. These frameworks are usually used to simulate and predict the behavior of large networks of neurons, from which neural mass models are often derived.
    \item Class II, or mesoscopic-level simulators, intends to simulate networks of interconnected neurons to simulate microscopic and mesoscopic neural circuits such as the internals of subcortical ganglia or cortical columns. There are other frameworks created with the specific objective of simulating mesoscopic systems, such as NENGO~\cite{Bekolay2014:NENGO}, focusing on applications in cognitive science, or Brain Modeling Toolkit~\cite{Dai2020:BrainModelingToolKit}, focusing on multiscale neural population circuits. Despite their obvious and many advantages, these software frameworks are most specifically designed for whole-brain modeling, so they are rarely used in this context given the difficult process of calibrating the models.
    \item Class III, or whole-brain simulators, aims at simulating whole-brain network activity at the resolution of interconnected brain regions. The main ingredients of these models are that a neural population model simulates each brain region (ROI), typically interconnected via the white matter fibers as captured by tract tracing or tractography based on diffusion imaging. One of the first and main open platforms for this endeavor is \emph{The Virtual Brain} (TVB)\footnote{\href{https://www.thevirtualbrain.org/}{https://www.thevirtualbrain.org/}}, which, since its inception, has gained wide recognition as the main platform for full-brain simulations, being a key element in the world’s first clinical trial on predictive brain modeling in epilepsy surgery~\cite{Ritter2013,SanzLeon2013}. Other options are PyRates~\cite{Gast2019}, and neurolib~\cite{Cakan2021}, which have been successfully applied to simulate a few use cases.
\end{itemize}

The Virtual Brain is the most extended and popular framework for Class III whole-brain modeling. It is, in essence, a neuroinformatics platform with a brain simulator that incorporates a range of neuronal models and dynamics at its core. It seamlessly integrates computational modeling and multimodal neuroimaging tools~\cite{Ritter2013,SanzLeon2013}, allowing the model-based simulation, analysis, and inference of neurophysiological mechanisms over several brain scales that underlie the generation of macroscopic neuroimaging signals. With TVB, creating personalized virtual brains is possible~\cite{Schirner2015,SanzLeon2015}, as well as studying the intricate multi-scale neural mechanisms within the brain~\cite{Schirner2018}. The most recent advances in TVB include its integration with cloud services through the European platform EBRAINS~\cite{Schirner2022}.

However, the original TVB implementation ---which we will refer to as \emph{TVB-Python} in the following--- has in computational speed its biggest drawback. To overcome slow execution speeds, two different strategies have been implemented. The first one, TVB-HPC\footnote{\href{https://wiki.ebrains.eu/bin/view/Collabs/rateml-tvb}{https://wiki.ebrains.eu/bin/view/Collabs/rateml-tvb}}~\cite{vanderVlag2022}, automatically produces high-performance codes for CPUs and GPUs using an easy XML-based language called RateML for model specification. RateML is based on the domain-independent language \emph{LEMS}~\cite{Vella2014}, which allows for the declarative description of computational models using a simple XML syntax. The already existing example implementations can be easily adapted to test different models, without requiring any knowledge about algorithmic optimization. The second one, Fast\_TVB (first used in \cite{Schirner2018}), is a specialized high-performance implementation of the "Reduced Wong Wang" model~\cite{Deco2014} written in C. It makes use of several optimization strategies and a sparse memory layout to efficiently use CPU resources, which makes it possible to simulate extremely large models with millions of nodes even on a standard computer in a reasonable time. Aquilu\'e-Llorens et al.~\cite{AquiluLlorens2023} presented a CPU-based version of the TVB-AdEx model~\cite{diVolo2019,Zerlaut2018}, used to study mechanisms underlying the emergence of conscious- and unconscious-like brain state dynamics using an MPI multi-node CPU setup. More recently, \cite{vanderVlag2023} presented the TVB-HPC framework, a modular set of methods used here to specifically implement the TVB-AdEx model for GPU and analyze emergent dynamics, accelerating simulations and thus reducing computational resource requirements, which in turn enabled larger parameter-space explorations.

However, despite its broad acceptance as \emph{the} reference platform for whole-brain simulations, it still exhibits some limitations when it comes to very large brain simulations, mainly in terms of its speed. For instance, a simulation with a parcellation with around 400 ROIs using the Balanced Excitation-Inhibition (BEI) model~\cite{Deco2014} using BOLD signals of around 10-minute recordings (approximately XX temporal samples, or volumes) takes around 5 minutes in a high-end processor. Suppose this evaluation is inserted into a fitting procedure where hundreds of evaluations take place. In that case, the total time for a single experiment on a single subject may rise to 1 week, which can become a serious problem even if using powerful parallel supercomputers.

\section{TVB Python}\label{sec:TVBPython}

\begin{figure}[ht]
 \centering
 \includegraphics[width=1.\linewidth,trim={0 15.0cm 0 0},clip]{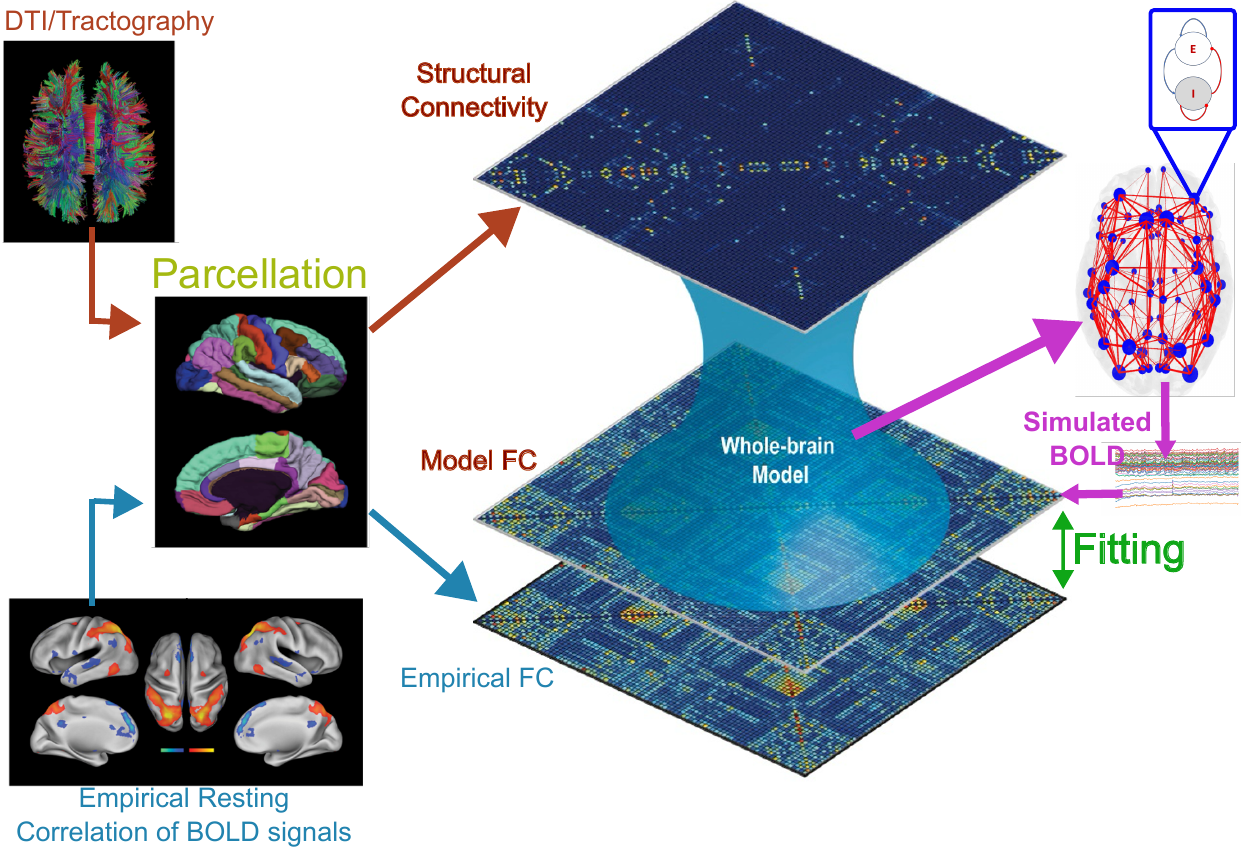}  
 \caption{Overview of the original TVB and the \fTVB{} system pipelines.}
 \label{fig:overview}
\end{figure}

The Virtual Brain (TVB) stands as a neuroinformatics software platform designed for whole-brain network simulations, leveraging biologically realistic connectivity to facilitate model-based inference of neurophysiological mechanisms across diverse brain scales. See Figure~\ref{fig:overview}. Through the incorporation of such connectivity, TVB enables the generation of macroscopic neuroimaging signals, including functional MRI (fMRI), EEG, and MEG, as depicted in Figure~\ref{fig:overview}. By harnessing individual subject data, TVB empowers the reproduction and assessment of personalized brain configurations. This personalized approach fosters exploration into the ramifications of pathological alterations within the system, thereby facilitating investigations into potential strategies to mitigate adverse processes. Within TVB, a biologically plausible, large-scale connectivity of brain regions is integrated. This connectivity is mediated through long-range neural fiber tracts identified via tractography-based methods. One differential trait of TVB and \fTVB{}, is their capability to handle connectivity both at the strength level, and also the time-delays associated to fiber lengths. In practice, the implementation for weights and tract lengths entails full matrices without any inherent symmetry constraints.

\begin{figure}[ht]
 \centering
 \includegraphics[width=1.1\linewidth]{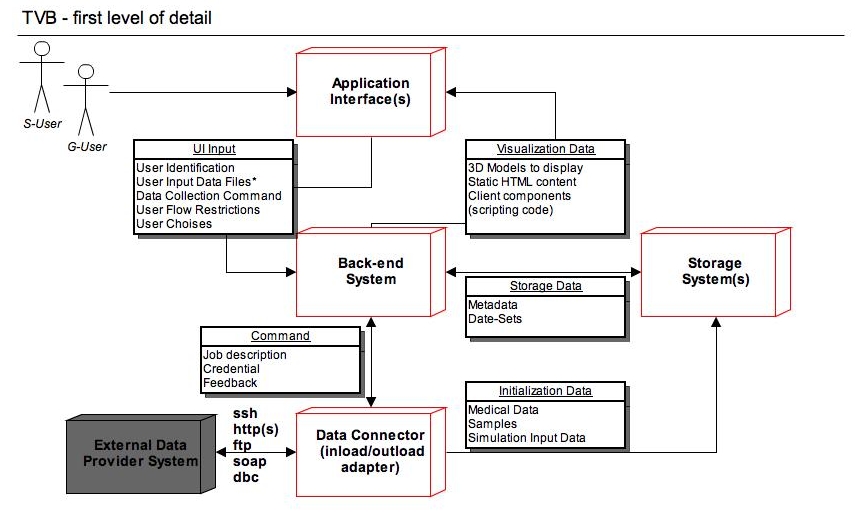}
 \caption{First level of detail of the architecture of the TVB system. Taken from the TVB main documentation.}
 \label{fig:TVB_architecture}
\end{figure}

From a high-level perspective, the architecture of TVB, as illustrated in Figure~\ref{fig:TVB_architecture}, accommodates interaction through various interfaces tailored to specific user needs and deployment scenarios. Structurally, TVB comprises two primary components: the backend, referred to as the "scientific computing core," and the supporting framework, which includes the graphical user interface and data visualization tools. TVB relies on an efficient and resilient storage data system to manage both metadata information and the underlying data. Functionally, TVB backends are designed to seamlessly support various interfaces, ensuring consistent functionality regardless of the selected interface.

In essence, a TVB application defines the logic necessary for data input, processing based on configurable parameters, and visualization of results. In the original Python implementation of TVB, while a single simulation step may not be readily distributable across multiple processing threads, each simulation step can be computed independently in different threads, potentially across distinct nodes. TVB documentation advises the deployment of complex simulation tasks on clusters to mitigate computational expenses.

As previously mentioned, TVB-Datatypes serve as a middleware layer facilitating the management and flow of data between the scientific kernel and the supporting framework. Essentially, TVB-Datatypes are annotated data structures encompassing one or more data attributes alongside associated descriptive information, inclusive of requisite methods for data manipulation.

The simulation core, the final module in our delineation, utilizes structural connectivity information within a comprehensive model of neural dynamics across the entire brain. Here, users can delineate the spatial structure upon which the model operates, as well as the hierarchical connections between nodes through the structural connectivity framework. Subsequently, the system numerically integrates the resultant coupled system of differential equations to replicate emergent brain dynamics. Central to this process, the Simulator class serves as both a repository for all pertinent information and as the conductor orchestrating the various steps associated with the simulation, all transparently to the user.

\section{TVB C++}\label{sec:TVB C++}

In response to the constrained performance of TVB Python, particularly evident when handling extensive networks or simulations necessitating extensive parameter exploration, \fTVB{} emerges as a simulator tailored for whole-brain network models (BNMs).  

\begin{figure}[ht]
 \centering
 \includegraphics[width=1.\linewidth,trim={0 20cm 0 0},clip]{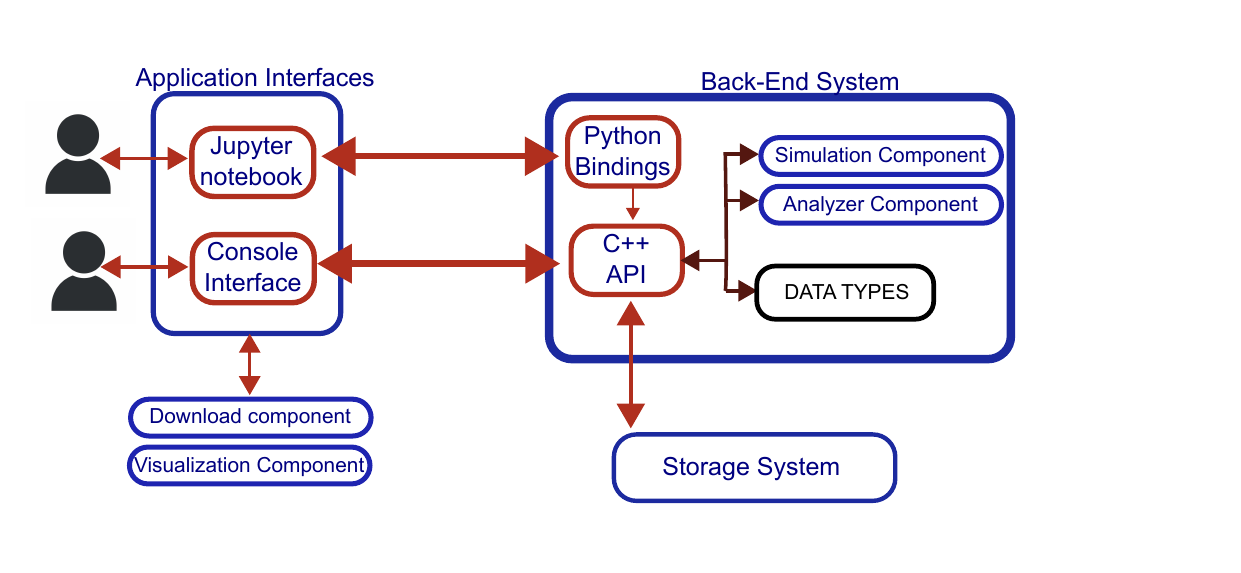}  
 \caption{Architecture of the \fTVB{} system. The architecture of \fTVB{} revolves around two distinct interfaces tailored to user interaction. The back-end consists of blocks that seamlessly interface with various top application layers. The \emph{data types} serve as the lingua franca between different components such as analyzers, visualizers, simulators, and uploaders. These datatypes embody "active data," meaning that when \fTVB{} is configured with a database, data encapsulated data type instances are automatically kept. Presently, the console interface operates independently of the storage layer, storing results solely in memory. Consequently, users employing the console interface must manually manage data import and export operations.}
 \label{fig:TVBC++_architecture}
\end{figure}

Reproducing the functionality of TVB-Python, \fTVB{} can simulate neural population models coupled by structural connectivity, which is an aggregated representation of the brain's white matter axon fiber bundle network. In addition to being implemented as a C++ library, \fTVB{} offers Python bindings to seamlessly plug it into existing pipelines, and subsequently serve as a fully integrated TVB Back-End. This way, by simply configuring the TVB simulator to use \fTVB{} instead of the default implementation in Python, existing code can be used without changes. 
The general architecture of \fTVB{} as a back-end for TVB is illustrated in Figure~\ref{fig:TVBC++_architecture}.

The architecture of \fTVB{} centers on two distinct interfaces tailored to user interaction. Currently, the console interface operates independently of the storage layer, storing results solely in memory. Consequently, users utilizing the console interface must manually oversee data import and export operations. On the other hand, its back-end comprises blocks that seamlessly interface with various top application layers. The \emph{data types} serve as the common language between different components such as analyzers, visualizers, simulators, and uploaders. These datatypes embody "active data," signifying that when \fTVB{} is configured with a database, instances encapsulating data types are automatically retained.

In the following subsections, we will introduce the main \fTVB{} components, namely its inputs, code structure, whole-brain models, integrators, history, coupling, and monitors. All these components represent functional modules that are completely user-configurable (see Section~\ref{sec:TVBPython}), allowing the users to interact with them to tailor their simulation and analysis needs.

\subsection{Inputs}

In general, \fTVB{} can load brain model data from a variety of sources and formats, with easy expandability to new formats and registration modalities. One of the main inputs for brain data consists of a structural connectome SC (in our case, weights files) and a parameter set (parameter file), plus spontaneous blood-oxygenation-level-dependent (BOLD) as stored in functional MRI data (fMRI), together with some regional information that depends on the problem at hand, from neuroreceptor data to gene expressions. Optionally, the user can provide information about the lengths of the tracts between the nodes, which would introduce \emph{delays} in the communications between the regions. This is also considered in the \fTVB{} implementation.

\subsubsection{Data formats}

While TVB allows loading data from Matlab data files (MAT), our version uses the Numpy (NPZ) format. We provide a Python script to convert MAT files into NPZ files.

\subsection{Code structure}

\begin{figure}[ht]
 \centering
 \includegraphics[width=1.0\linewidth]{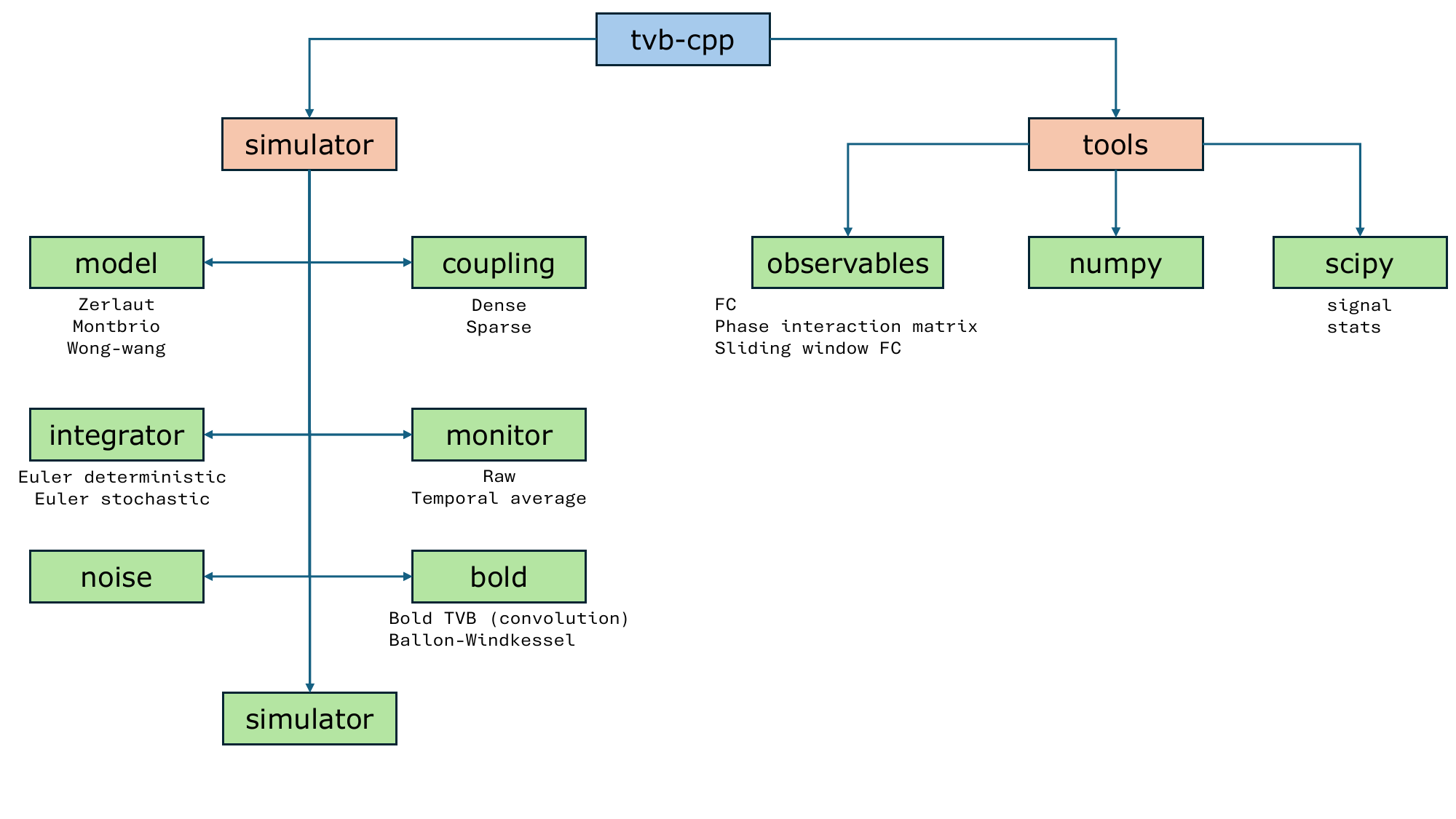}
 \caption{Main modules of \fTVB{}. On the left branch are all the core simulation modules, i.e., model, coupling, integrator, monitor, noise, BOLD, and the simulator itself. On the right, are the tools, e.g., observables, and the numpy and scipy interfaces.}
 \label{fig:module-diagram}
\end{figure}

\fTVB{} replicates the same structure of the original TVB library. That means that the structure of the classes is the same, providing the same functionality and flexibility. However, \fTVB{} only focuses on the simulation part of the original TVB, and it does not provide any user interface. The two main parts of the library are the simulator module {\it modulename.py}, consisting of the population models to simulate the activity of one brain region, e.g., mean-field AdEx, Wong-Wang, FRE; coupling, which can be either dense or sparse; the integrator, where the different integration schemes are implemented; the monitors, including raw data monitor and the temporal averaging monitor, as well as a convolution-based BOLD; the noise implementation; the BOLD conversion system based on the Ballon-Windkessel model; and the simulator module itself, in charge of wrapping all the other components in a single, uniform shell. On the other hand, the other big module, the tools, is composed of a series of submodules, being the observables, e.g., FC, FCD, and the Phase Interaction Matrices a few examples, as well as the interfaces with numpy and Scipy. See Figure~\ref{fig:module-diagram}.

While TVB is heavily based on NumPy, no version of this library exists in C++. To provide similar functionality we have chosen the \emph{Eigen} library~\cite{eigenweb}, specifically version 3.9 since it provides advanced matrix indexing similar to NumPy. Moreover, Eigen takes advantage of SIMD instruction sets, and it can be combined with BLAS/LAPACK, Intel MKL library, and CUDA.

All the objects that can be configured in a simulation are:
\begin{itemize}
    \item Whole-Brain Models
    \item Integrators
    \item History
    \item Coupling
    \item Monitors
\end{itemize}

\subsection{Population and mean-field models of regional activity}

\fTVB{} has a modular design that allows the incorporation of a large variety of whole-brain models. Model definitions are very similar to TVB, and with \fTVB{} this implies deriving from the abstract class \texttt{tvb::Model}. The main method to implement is \texttt{tvb::Model::operator()} (equivalent to the \texttt{dfun()} method in TVB) that computes the derivatives of the state variables.

Currently, several methods have been implemented and tested, namely, the Balanced Excitation-Inhibition (BEI) model by Deco et al.~\cite{Deco2014}, the Montbri{\'{o}}, Paz{\'{o}} and Roxin's FRE model~\cite{Montbri2015,Montbrio_2020}, and the mean-field AdEx~\cite{diVolo2019,Boustani2019,Carlu2020,Goldman2023}. See Section~\ref{sec:useCases}.

\subsection{Numerical integrators}

TVB allows the use of different integrators to compute simulations with the same model. \fTVB{} follows the same approach and provides an abstract class \texttt{Integrator} that can be derived and this way implements different integration strategies. Right now we provide the Euler deterministic and stochastic (Euler-Maruyama) integrators. In this latter case, stochastic integrators can be combined with different noise generators, in the same way, that TVB does.

\subsection{History and time-delays}

TVB takes into account the delay in signal transmission between nodes of the brain model. The delays are computed from the tract length matrix and the signal speed. \fTVB{} also keeps track of the signal history and provides two modes: a dense history that keeps track of all signals at all possible delays, and a no-delay history, which is more efficient and can be used in some models.

\subsection{Coupling}

The coupling functions allow activity (state variables) that have been propagated over the long-range connectivity to pass through these functions before entering the equations (\texttt{tvb::Model::operator()}) describing the local dynamics.

At the moment, only linear coupling is implemented, but other functions are straightforward to use.

\subsection{Monitors}

Another feature of TVB is the Monitors, objects that record significant values from the simulation. In their simplest form, they return all the simulated data, subsampled data, spatially averaged temporally subsampled, or temporally
averaged subsamples. The current version of \fTVB{} implements the two former ones.

TVB also has more elaborate monitors that instantiate a physically realistic measurement process on the simulation, such as EEG, MEG, and fMRI (BOLD). In \fTVB{} we follow a slightly different approach and these measurement processes are performed after the simulation, using the monitor data as input.

\subsection{Extra tools}

To make \fTVB{} a useful tool, the ability to simulate Whole-Brain models is just the first step. There are many extra tools needed to analyze the results of such simulations. \fTVB{} provides some of these tools, and provides an example implementing all the necessary steps to reproduce the main results of the corresponding models enumerated above and described in Section~\ref{sec:useCases}: The BEI model~\cite{Deco2014}, the FRE model~\cite{Montbri2015,Montbrio_2020} and the mean-field AdEx model~\cite{diVolo2019,Boustani2019,Carlu2020,Goldman2023}. We have chosen this work since it allows us to compare standard features already existing in TVB and high-performance implementations. To reproduce these works, some steps need to be taken:
\begin{itemize}
    \item Perform simulations using the corresponding model
    \item Estimate the BOLD signal from the simulated physiological signals.
    \item Compute some sort of observable, e.g., sliding-window correlations, from the obtained BOLD signals. See Section~\ref{sec:observables}.
    \item Use a measure, e.g., the $L^2$ metric, between the empirical and simulated observables obtained.
\end{itemize}

For the last two items, we had to implement some of the features of NumPy and SciPy into \fTVB{}, like the Kolmogorov statistic.

\subsubsection{Estimate BOLD signals from neural output}

Once we have obtained the simulated mean field activity, we need to transform it into a Blood Oxygenation Level Dependent (BOLD) signal. Given that the conversion from electrical activity to a BOLD signal is, in its mathematical sense, a low-pass filter, this process can be computed in two different ways: as a convolution with Volterra kernels, or as a generalized hemodynamic model such as the Balloon-Windkessel model. In the following, we describe both implementations, as \fTVB{} implements the two methods to create a BOLD signal from a raw signal produced by the simulation, as the original TVB implementation also does.

\paragraph{Haemodynamic Response Function Model} 
One option is to use convolutions with Hemodynamic Response Function (HRF) to compute the output BOLD signal, derived using generic techniques from nonlinear system identification~\cite{Buxton1997,Friston2000,Boynton1996,Polonsky2000,Glover1999}. The approach employed allows for the estimation of Volterra kernels, which elucidate the relationship between stimulus presentation and consequent hemodynamic responses. This technique represents the BOLD signal as the outcome of a convolution with a Volterra series~\cite{Friston1998}, which are, essentially, Taylor expansions extended to encompass dynamical input-state-output systems and consider the impact of the input at all times in the recent past~\cite{Friston2000}. As a result, they are regarded as model-independent, capable of representing any such dynamical system. Serving as high-order extensions of linear convolution or "smoothing," Volterra kernels provide a nonlinear characterization of the hemodynamic response function. Consequently, they facilitate the modeling of responses to stimuli across various contexts (e.g., different rates of word presentation) and interactions among stimuli.
In \fTVB{} this is implemented as a special case of a \emph{Monitor} class, which are tools to record significant values from the simulation, in this case using HRF kernels. As we are following TVB's main implementation, we are also using the first-order Volterra kernel, too.

\paragraph{Balloon-Windkessel Model:} 
We implemented the generalized hemodynamic model of Stephan et al.~\cite{Obata2004,Stephan2007}. We compute the BOLD signal in the $k$-th brain area from the firing rate of the excitatory pools $H^{(E)}$, such that an increase in the firing rate causes an increase in a vasodilatory signal, $s_k$, that is subject to auto-regulatory feedback. Blood inflow $f_k$ responds in proportion to this signal inducing changes in blood volume $v_k$ and deoxyhemoglobin content $q_k$. The equations relating to these biophysical variables are:

\begin{equation}
\begin{split}
\frac{d s_k}{dt} &= 0.5 r_k^{(E)} + 3 - k s_k - \gamma (f_k-1) \\
\frac{d f_k}{dt} &= s_k \\
\tau \frac{d v_k}{dt} &= f_k - v_k^{\alpha^{-1}} \\
\tau \frac{d q_k}{dt} &= f_k \frac{1-(1-\rho)^{f_k^{-1}}}{\rho} - q_k \frac{v_k^{\alpha^{-1}}}{v_k}
\end{split}
\end{equation}
with finally
\begin{equation*}
B_k = v_0 \left[ k_1(1-q_k)+k_2(1-\frac{q_k}{v_k})+k_3(1-v_k) \right]
\end{equation*}
being the final measured BOLD signal.

We used the updated version~\cite{Stephan2008} described later, which consists of introducing the change of variables $\hat{z} = ln z$, which induces the following change for $z=f_k$, $v_k$ and $q_k$, with its corresponding state equation $\frac{dz}{dt} = F(z)$, as:
\begin{equation*}
\frac{d\hat{z}}{dt} = \frac{d\;ln(z)}{dz} \frac{dz}{dt} = \frac{F(z)}{z}
\end{equation*}
which results in $z(t)=exp(\hat{z}(t))$ always being positive, which guarantees proper support for these non-negative states, and thus numerical stability when evaluating the state equations during evaluation.

\subsubsection{Observables}
\label{sec:observables}

\paragraph{Functional Connectivity:} 
It is computed as the Pearson correlation between the model and empirical estimates of static (i.e., time-averaged) Functional Connectivity, estimated across all pairs of brain regions (FC); more specifically, the correlation between the values in the upper triangles of the model and empirical FC matrices.

\paragraph{Sliding-Window Functional Connectivity Dynamics (swFCD):} 
In~\cite{DECO20183065}, to take into account the spatiotemporal fluctuations in functional brain dynamics over time, the model is fitted to the spatiotemporal dynamics of the data (i.e., to the Functional Connectivity Dynamics [FCD])~\cite{Hansen2015525, Allen2012, Deco2016125}. The measure used is a sliding-window correlation between BOLD signals.
To implement this, \fTVB{} re-implements some functions from NumPy and SciPy.

\paragraph{Phase Functional Connectivity Dynamics (phFCD):} 
From each BOLD time series of each ROI, we can Hilbert-transform them to yield the phase evolution of the regional signals, which can be averaged, to finally be used to compute the KS distance from other identically processed series. This yields the Phase Functional Connectivity Dynamics (phFCD), which is a 3D matrix of $N\times N\times T$-size, where $N$ is the number of nodes in the parcellation used and $T$ indicates the number of image volumes acquired in all considered sessions.

\subsection{Python bindings}

Compiling and executing C++ code can be challenging to most neuroscientists who are more used to user-friendly interfaces. To provide an easy entry point to use the power of \fTVB{}, we have developed a Python package that provides easy access to the features of this library. This package provides bindings for the different library components, allowing the selection of different models (e.g., FRE, Dynamic Mean Field, mean-field AdEx), different monitors (e.g., two different BOLD monitors), integrators (e.g., Euler or Euler-Maruyama) and observables (e.g., FC, swFCD, phFCD). The API supports interaction with the original TVB simulator python class so that it can be used as an alternative Back-End without the need for code adaptation.

The library is hosted in a Github public repository~\cite{Martin_TVB_C_Library_2022} and there is a folder that contains the Python binding including a \emph{Jupyter notebook}\footnote{\href{https://github.com/neich/tvb-root-cpp/blob/main/python/tvbcpp.ipynb}{https://github.com/neich/tvb-root-cpp/blob/main/python/tvbcpp.ipynb}} that makes it easier to test the library. The API as of now is straightforward and it allows to select of different neural population models, monitors, integrators, and observables, following the protocols and standards of the original TVB implementation, thus allowing for seamless integration as a backend for  TVB-python.

In Listing~\ref{code:PythonBindings}, a comprehensive example of \fTVB{} Python bindings is provided. It is essential to note that the \fTVB{} module operates as a state machine, requiring the configuration of all state variables before execution, starting with the integration variables and loading the structural connectivity matrix (lines 6-14). 
Subsequently, a neuronal mass model, FRE model with 6 state variables, is configured, with noise added only to the last two via the Euler-Maruyama (referred to as \emph{EulerStochastic}) integrator (lines 16-21). The initial state of the model's variables is set to zero (lines 22-25), followed by the configuration of global coupling (line 27).
While the current implementation only offers linear coupling, the library is designed for easy extension. Users can query model parameters before execution by utilizing the \texttt{get\_model\_parameters()} command (line 29), and change them before execution.
An example of this can be seen below, in Section~\ref{sec:montbrio}, where an example of a change of some complex model parameters is presented.

A monitor is then added to process the raw simulation data, averaging over 1 ms of state variable 0 (line 32). Finally, the simulation is run with the \texttt{run\_sim} command, specifying the time interval in milliseconds (line 34). The simulation output comprises a list of monitor results, with each result being a tuple containing two arrays: a 1D array with time samples and a 2D array with simulation output data, such as time-series data.

To convert the simulation output to a BOLD signal, three methods are available: \texttt{Stephan2007}, \texttt{Stephan2007b}, and the original \texttt{TVB} convolution-based monitor (lines 41-43). Users can select their desired monitor by changing the appropriate keyword.

Finally, the resulting signal is subsampled for illustrative purposes, and the results can be plotted using the code snippet provided in Listing~\ref{code:PythonPlotResults}.

\section{Use cases}
\label{sec:useCases}

In this section we provide practical use cases implemented within \fTVB{}. First, we present a modified version of the Balanced Excitation-Inhibition model~\cite{Deco2014}, followed by the implementation of the Mean Field Adaptive Integrate and Fire Neurons (mean-field AdEx) in its first order~\cite{Zerlaut2018} and second order~\cite{diVolo2019} variants. Finally, the FRE model~\cite{Montbri2015,Montbrio_2020} is described.

We have tested the \fTVB{} version against TVB-Python version 2.8.1, the FAST\_TVB implementation, and RateML.
The comparison involves two main tasks, the ones involving whole-brain simulations using a neural model, and other computations that use mainly external functions, like BOLD signal generation. It must be taken into account that FAST\_TVB only offers a reduced set of the full TVB functionality, so we only compared the functionality that was possible for us to compare. All comparisons were made in a computed with an Intel i9-13900H CPU (14 cores) and an NVidia laptop RTX 4070 GPU.

\subsection{Modified Balanced Excitation-Inhibition}
In this section, we will use \fTVB{} to reproduce the results previously obtained by Deco et al.~\cite{DECO20183065}. The full Balanced Excitation/Inhibition (BEI) model~\cite{Deco2014} at the whole-brain level is expressed by the following system of coupled differential equations:
\begin{equation*}
\begin{split}
I_k^{(E)} &= W_E\,I_o + w_+\,J_N \, S_k^{(E)} + J_N G \sum_j C_{kj} S_j^{(E)} - J_k S_k^{(I)}+ I_{ext}, \\
I_k^{(I)} &= W_I\,I_o + J_N S_k^{(E)} - S_k^{(I)} + \lambda J_N G \sum_j C_{kj} S_j^{(E)}, \\
r_k^{(E)} &= H^{(E)}(I_k^{(E)}) = \dfrac{M_k^E (a_E I_k^{(E)} - b_E)}{1 - \exp(-d_E M_k^E (a_E I_k^{(E)} -b_E))}, \\
r_k^{(I)} &= H^{(I)}(I_k^{(I)}) = \dfrac{M_k^I (a_I I_k^{(I)} - b_I)}{1 - \exp(-d_I M_k^I (a_I I_k^{(I)} -b_I))}, \\
\dot{S}_k^{(E)}    &= -\dfrac{S_k^{(E)}}{\tau_E} + (1 - S_k^{(E)}) \,  \gamma H^{(E)}(I_k^{(E)}), \\
\dot{S}_k^{(I)}    &= -\dfrac{S_k^{(I)}}{\tau_I} + H^{(I)}(I_k^{(I)})
\end{split}
\end{equation*}
Here, the last \emph{two} equations should add, when integrating, an uncorrelated standard Gaussian noise term with an amplitude of $\sigma = 0.01nA$ (Euler-Maruyama integration).

The parameters of this model were originally defined in Wong and Wang~\cite{Wong2006} to emulate resting-state conditions, such that each isolated node displays the typical noisy spontaneous activity with low firing rate ($H^{(E)}\sim 3Hz$) observed in electrophysiology experiments. Moreover, following Deco et al.~\cite{Deco2014}, the inhibition weight, $J_n$, was adjusted separately for each node $n$ such that the firing rate of the excitatory pools $H^{(E)}$ remains clamped at 3Hz even when receiving excitatory input from connected areas. 
It has been demonstrated that this mechanism leads to a better prediction of the resting-state FC and a more realistic evoked activity~\cite{Deco2014}. We refer to this model as the
balanced excitation-inhibition (BEI) model.

To study the effects of LSD, Deco and co-authors~\cite{DECO20183065} used local regional information, introduced in the form of a detailed map of $5-HT_{2A}$ receptor density of the neuromodulator serotonin, obtained from a new high-resolution human brain in vivo atlas. For this, they defined a global gain scaling parameter, added to the original fixed gain parameters (which were the same for all regions), and thus scaling the regional $5-HT_{2A}$ receptor values influencing the recursive circuits of excitatory and inhibitory neurons. In mathematical terms, we modify the value of $M_k^E$ in the above equations as
\begin{equation*}
M_k^E = g_E (1 + s \;\; 5HT2A_k)
\end{equation*}
with $g_E$ the value previously used~\cite{Deco2014}, $5HT2A_k$ the neuroreceptor density for the $k$-th region, and $s$ the scaling factor.
The procedure follows the somewhat standard pipeline in these cases: first, fit the model to the placebo condition but not the LSD condition, i.e., assuming zero values of the scaling that correspond to the original gain values. Then, a scaling value that fits the LSD condition (using the sensitivity of the functional connectivity dynamics) while still using the original whole-brain placebo model, but now including the new element of receptor binding, which modulates each region with the different empirical measures of the receptor binding. Thus, here we verify their results with \fTVB{}, showing that neurotransmitter modulation of whole-brain activity dynamics can be quantitatively ascribed to one type of receptor binding (here $5-HT_{2A}$) that modulates brain-wide neural responses.

\begin{figure}[ht!]
\includegraphics[width=0.3\textwidth,trim={0.7cm 0 1cm 0},clip]{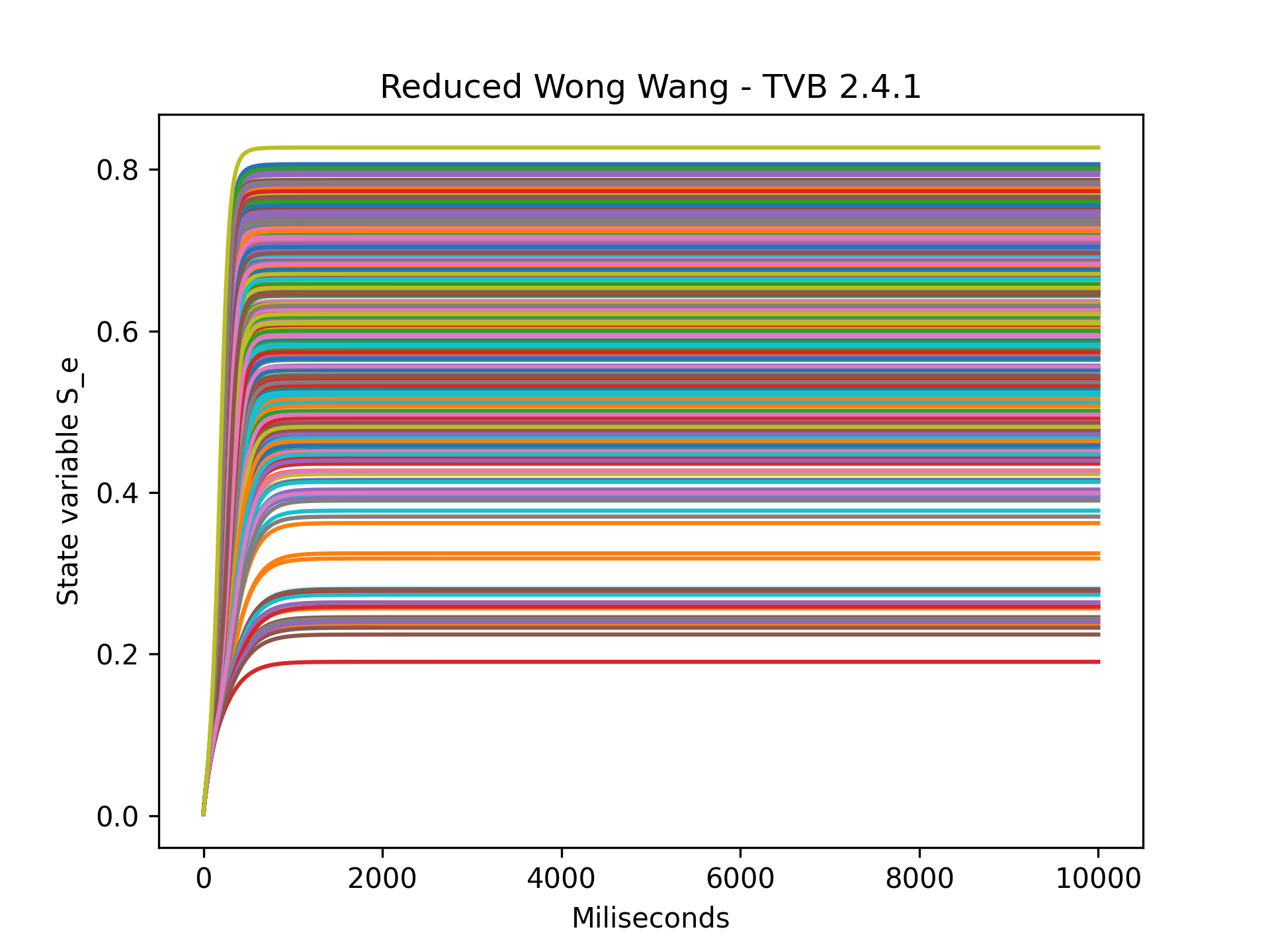}
\includegraphics[width=0.3\textwidth,trim={0.7cm 0 1cm 0},clip]{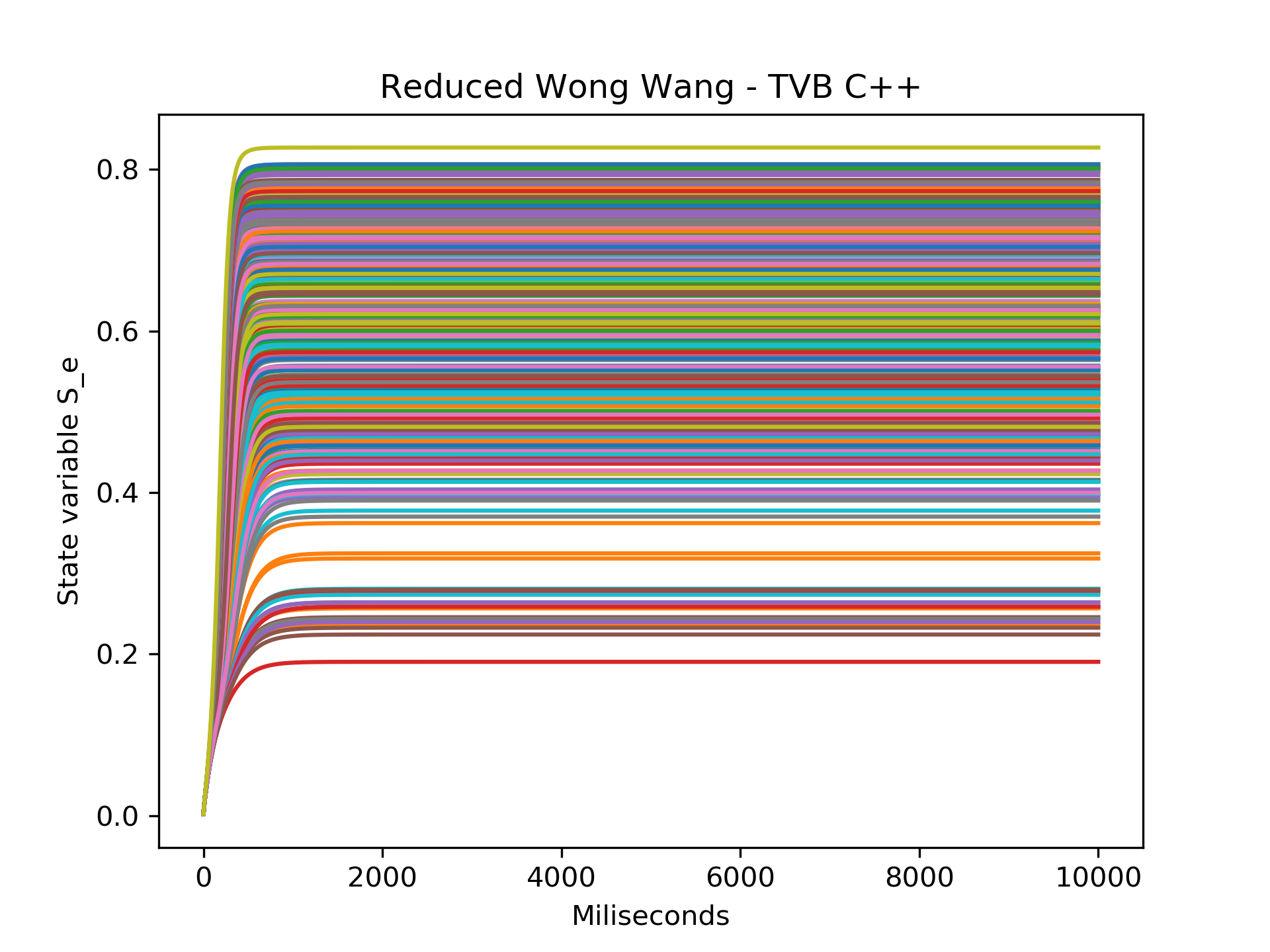} 
\includegraphics[width=0.3\textwidth,trim={0.7cm 0 1cm 0},clip]{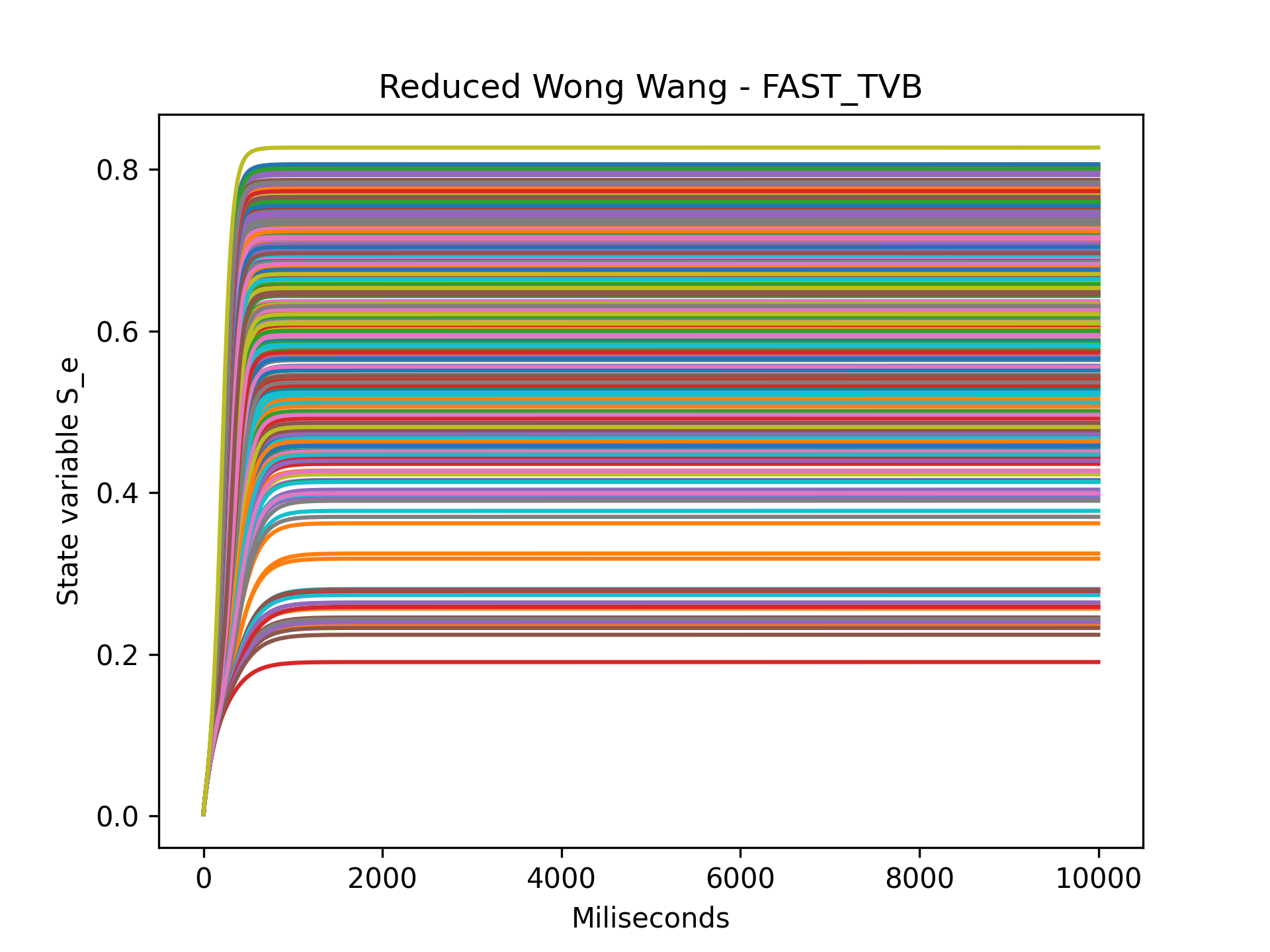} \\
\includegraphics[width=0.3\textwidth,trim={0.7cm 0 1cm 0},clip]{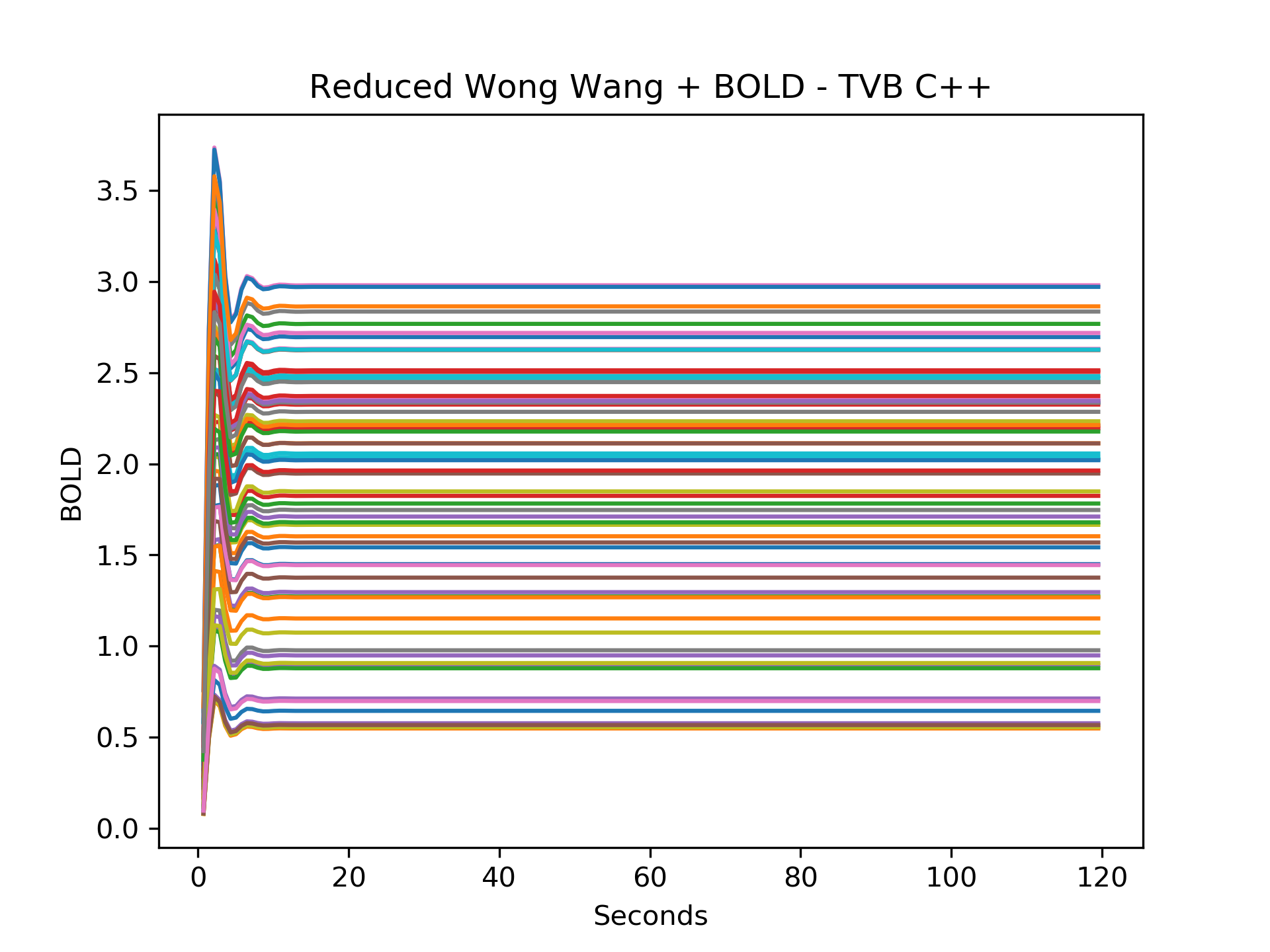}
\includegraphics[width=0.3\textwidth,trim={0.7cm 0 1cm 0},clip]{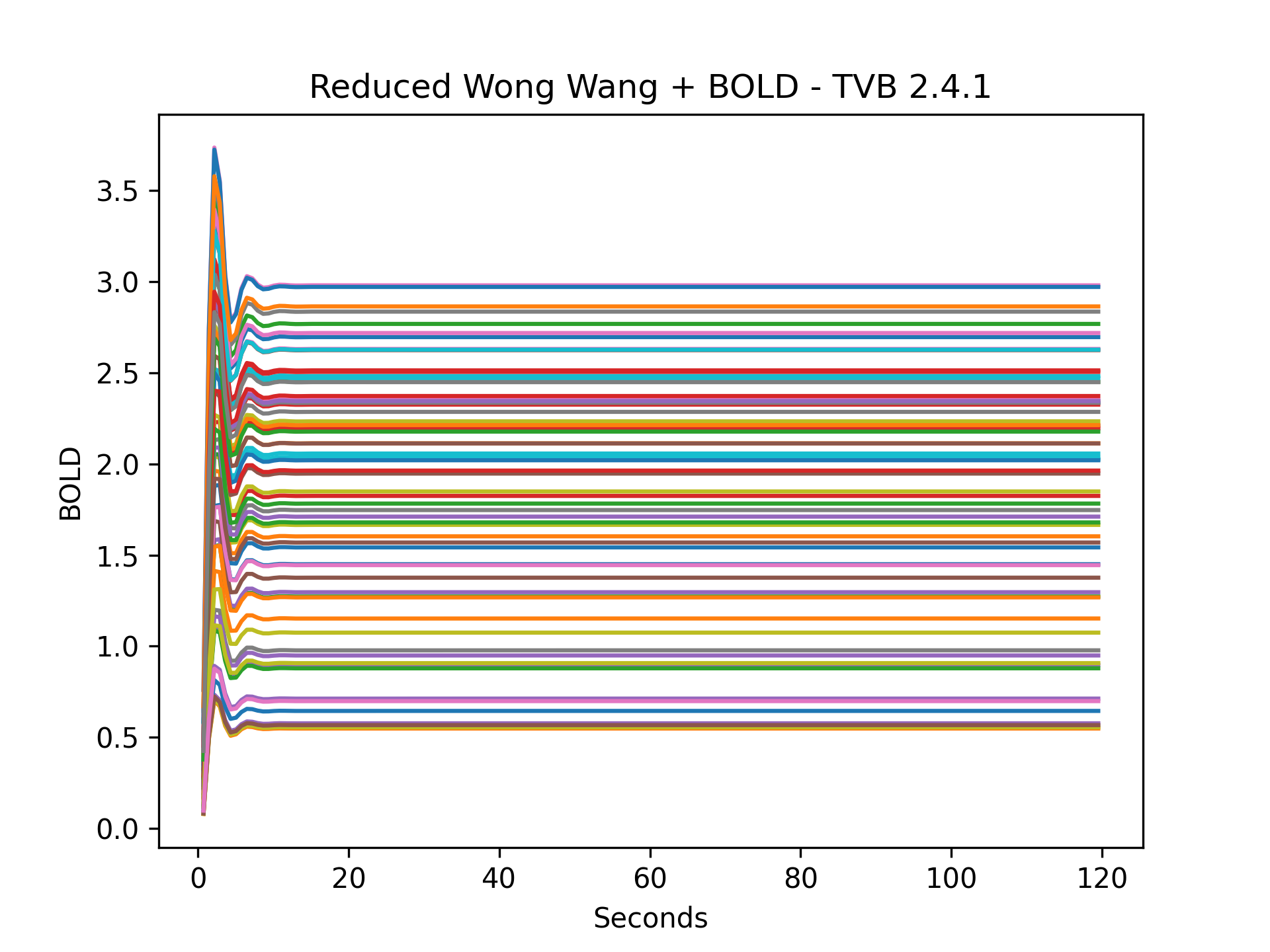} \\
\includegraphics[width=0.3\textwidth,trim={0.7cm 0 1cm 0},clip]{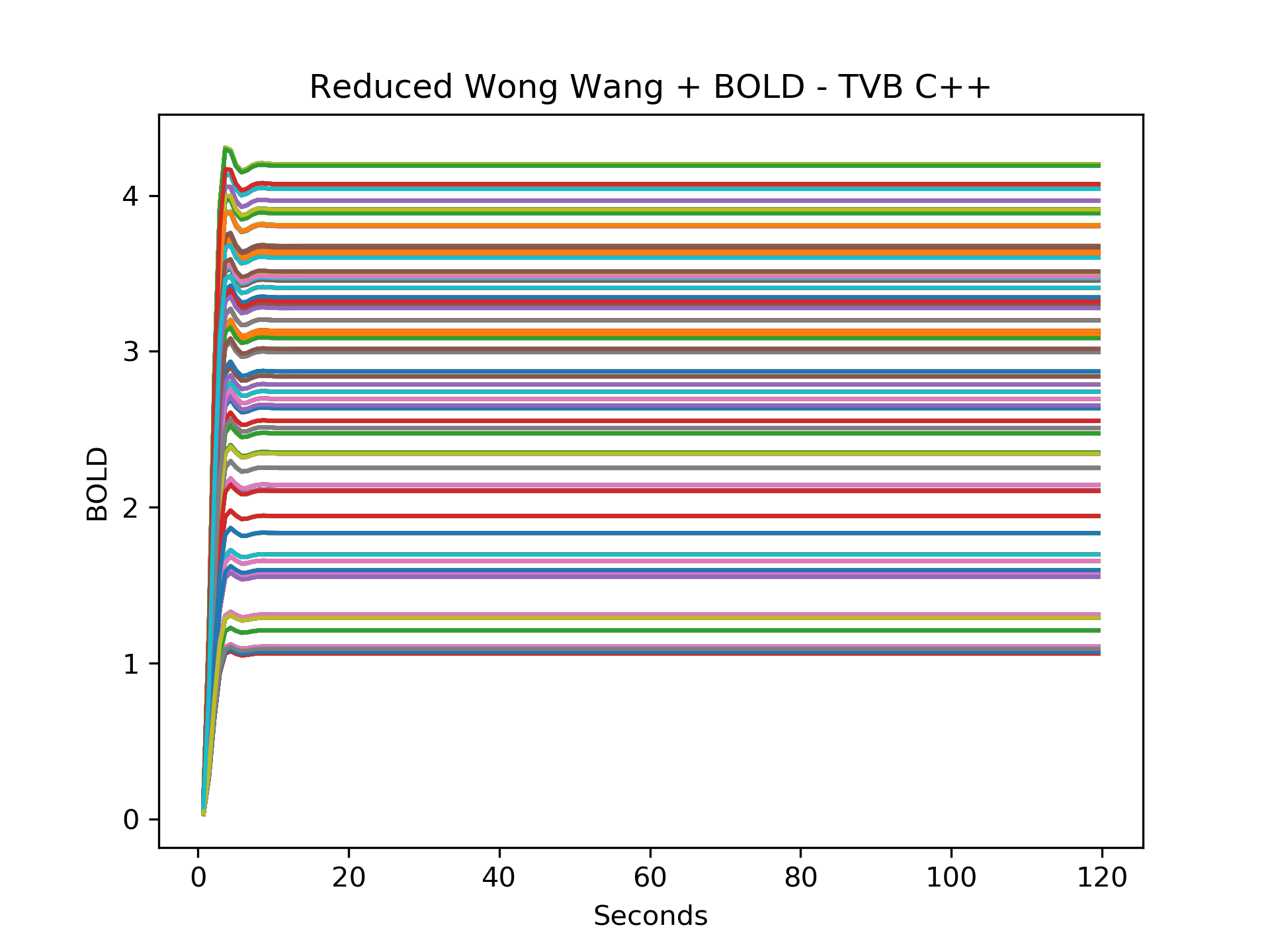}
\includegraphics[width=0.3\textwidth,trim={0.7cm 0 1cm 0},clip]{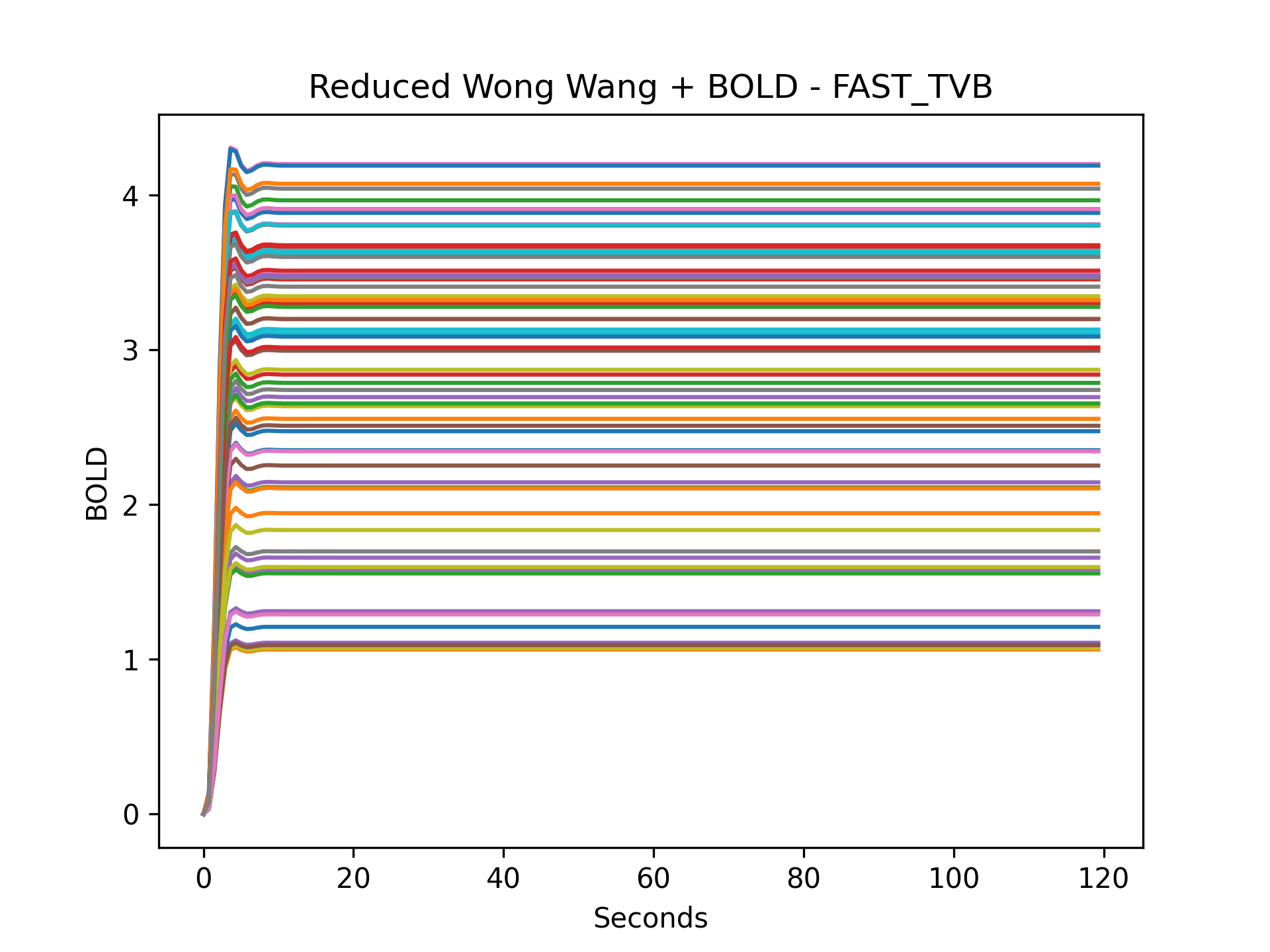}
\includegraphics[width=0.3\textwidth,trim={0.6cm 0 1cm 0},clip]{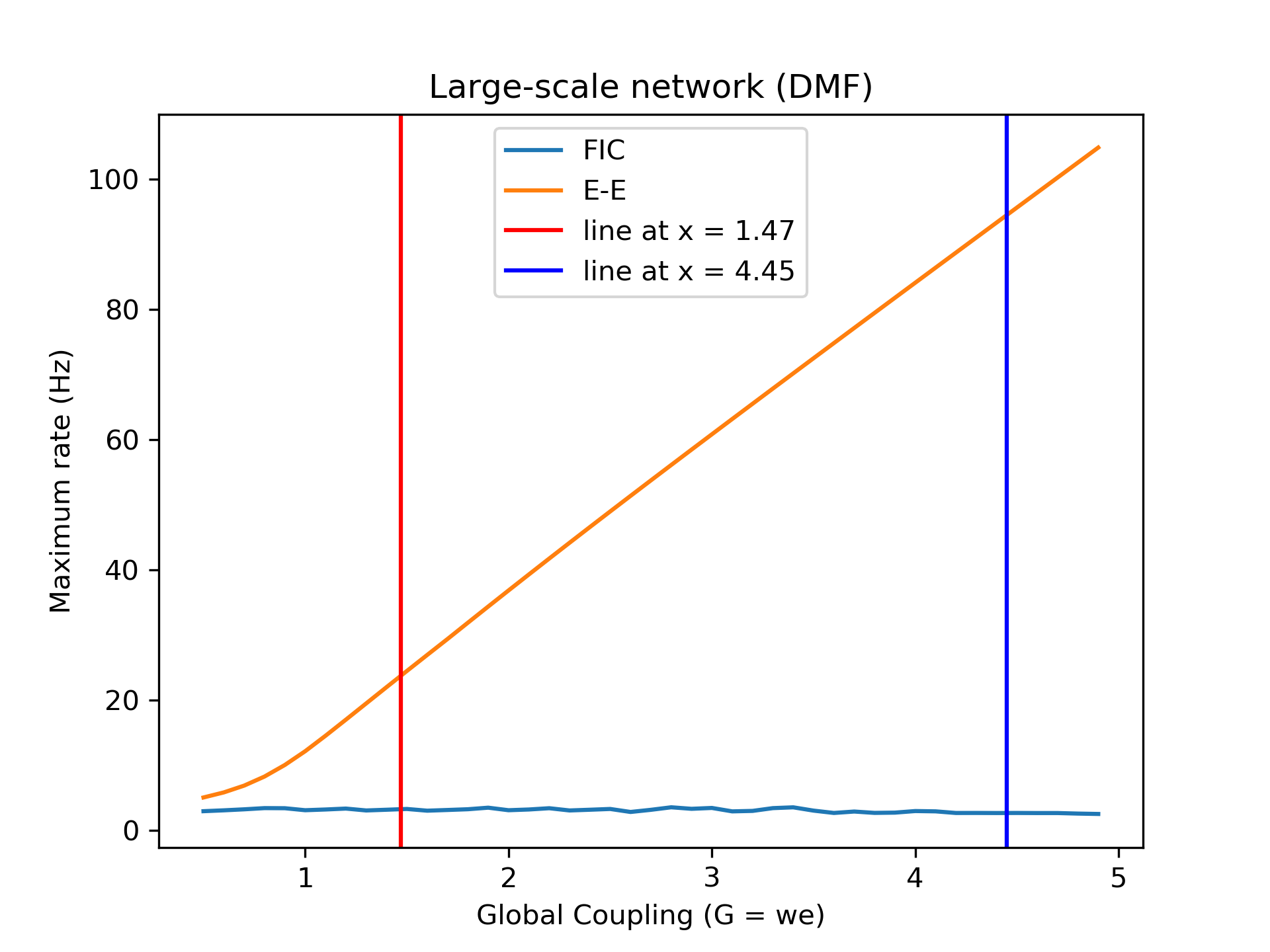}
\caption{Different plots comparing different platforms with the Reduced Wong Wang model.
Top: Plot of a (direct) simulation with TVB, FAST\_TVB, and \fTVB{}.
Middle: Plot of a BOLD signal with \fTVB{} and TVB, using HRF convolutions.
Bottom, first two columns: Plot of a BOLD signal from a whole brain simulation with \fTVB{} and TVB, using the Balloon-Windkesssel model. 
Observe that in all cases, all platforms compared in the same rows produce the exact same result.
Bottom right: Plot of the maximum firing rate of the neuron population for the same model, with and without the FIC mechanism.}
\label{fig:plotBold}
\end{figure}

\begin{figure}[ht!]
\includegraphics[width=0.3\textwidth,trim={0.7cm 0 1cm 0},clip]{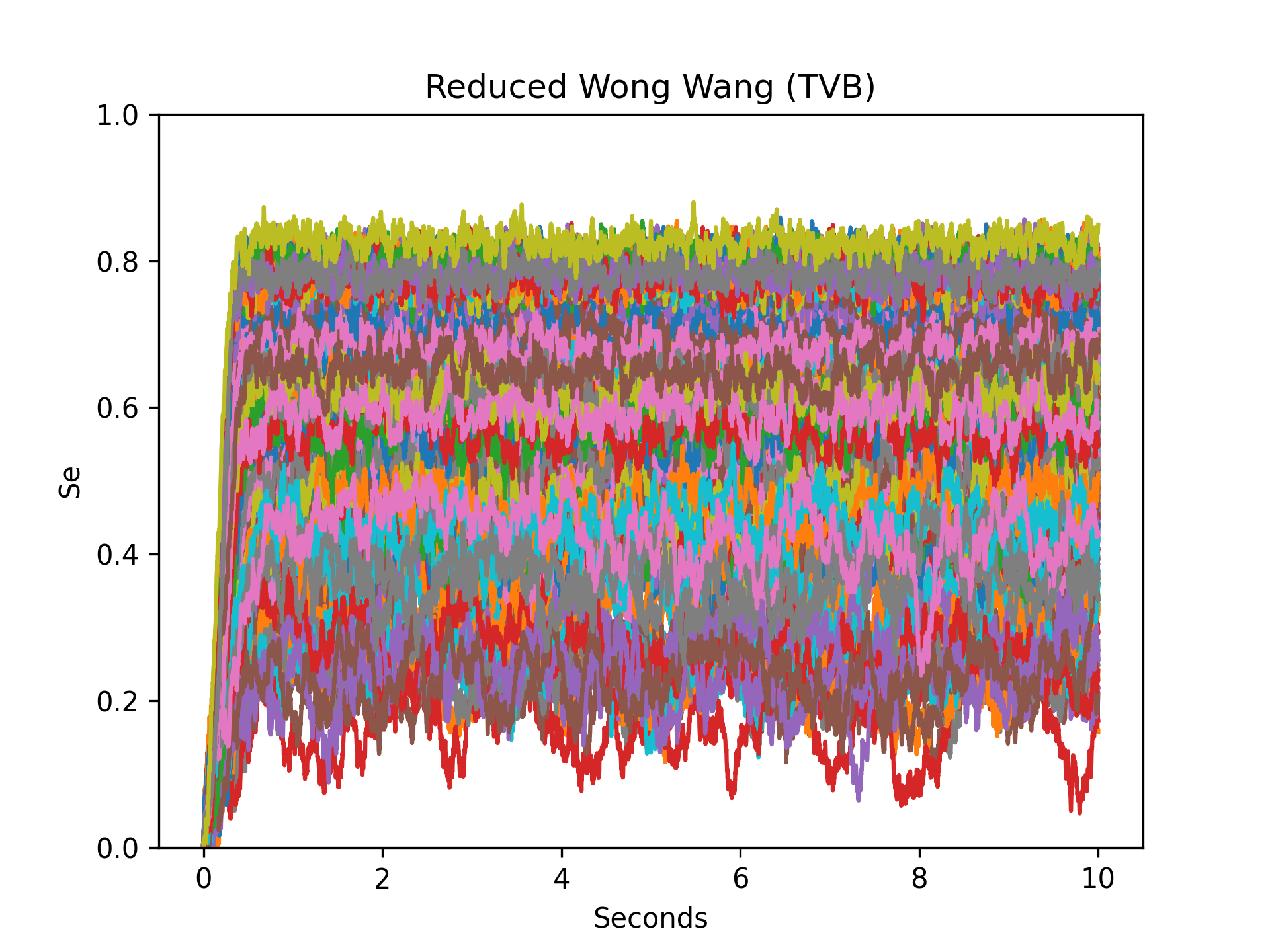}
\includegraphics[width=0.3\textwidth,trim={0.7cm 0 1cm 0},clip]{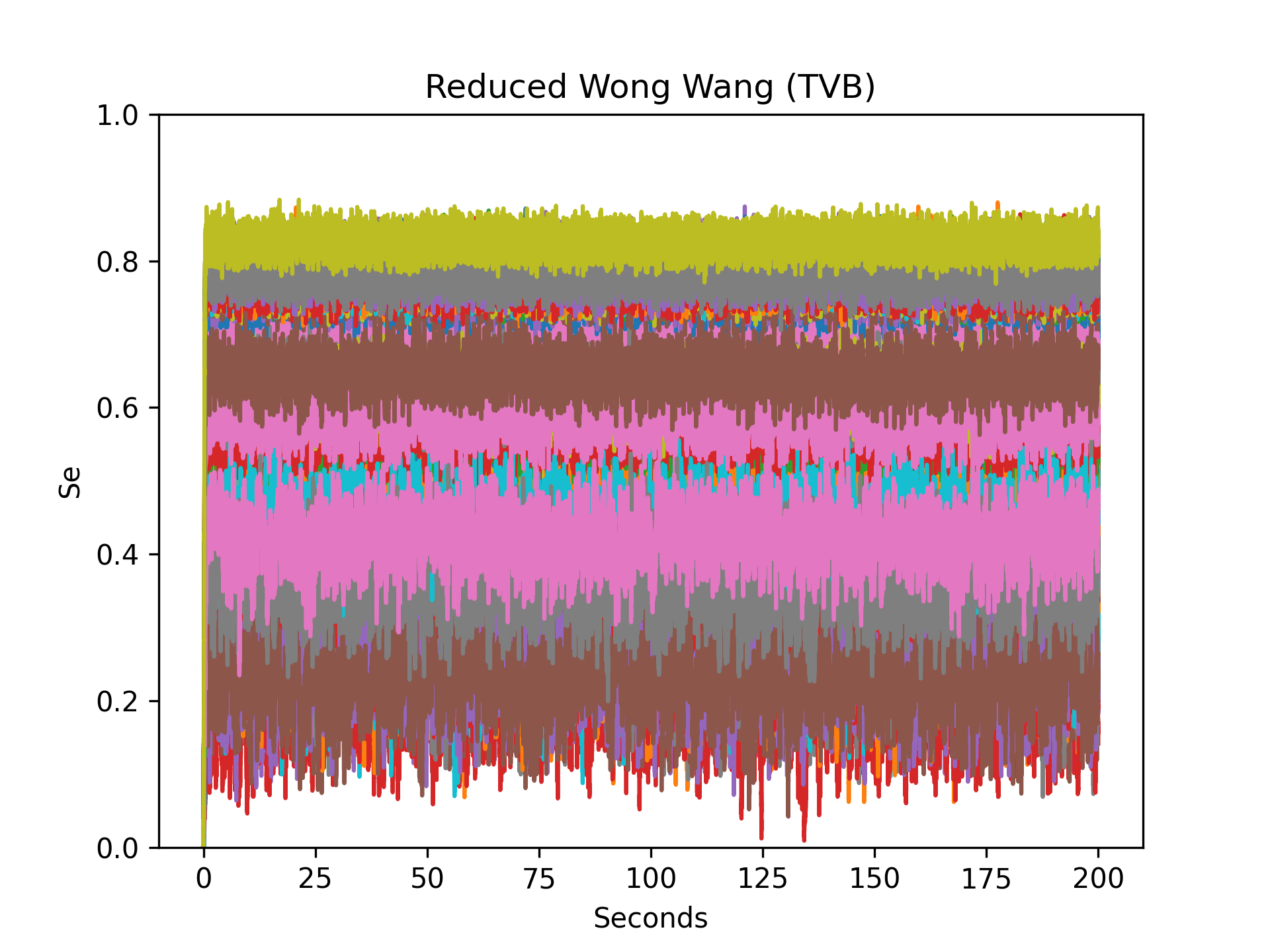} 
\includegraphics[width=0.3\textwidth,trim={0.7cm 0 1cm 0},clip]{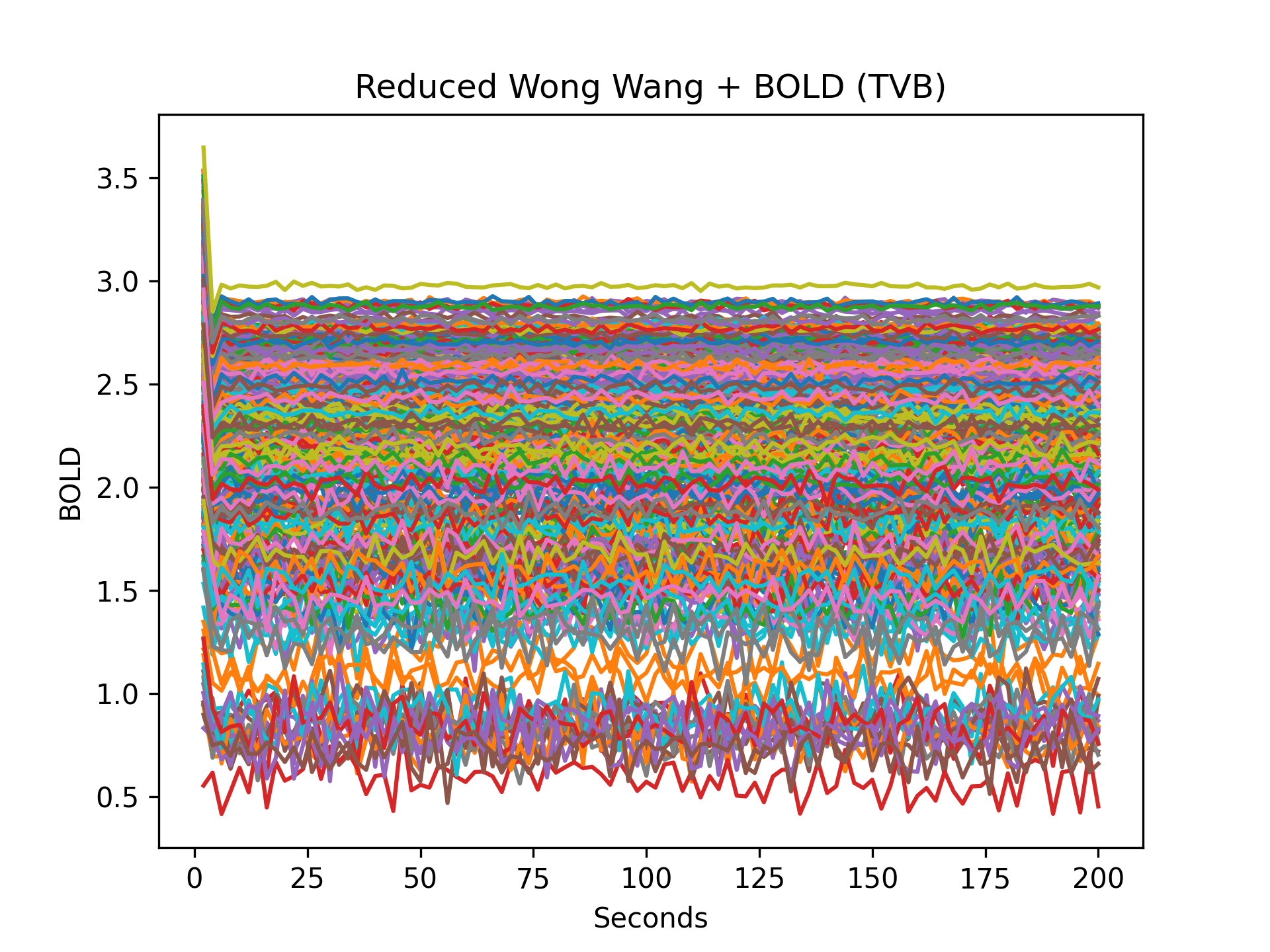} \\
\includegraphics[width=0.3\textwidth,trim={0.7cm 0 1cm 0},clip]{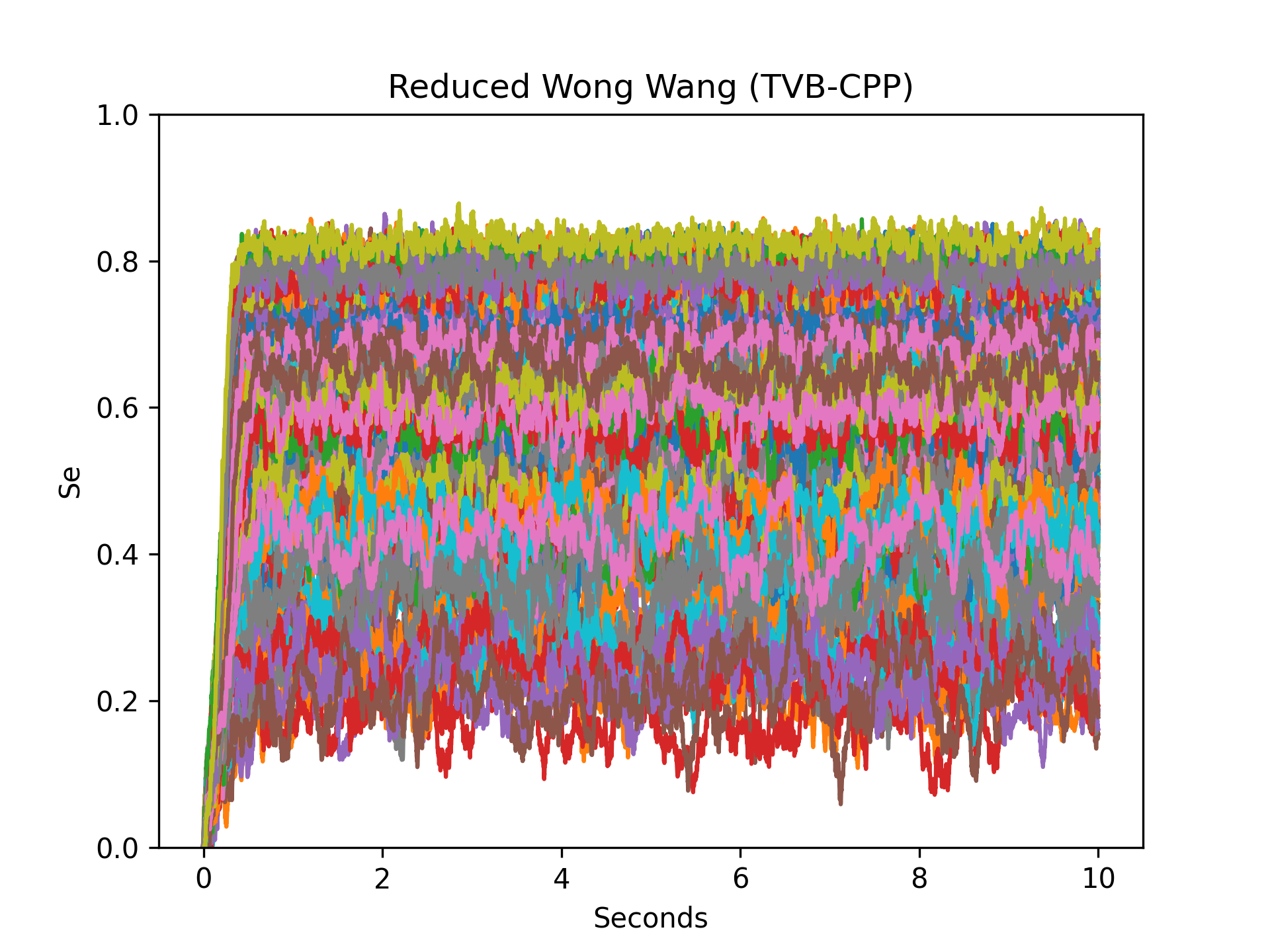}
\includegraphics[width=0.3\textwidth,trim={0.7cm 0 1cm 0},clip]{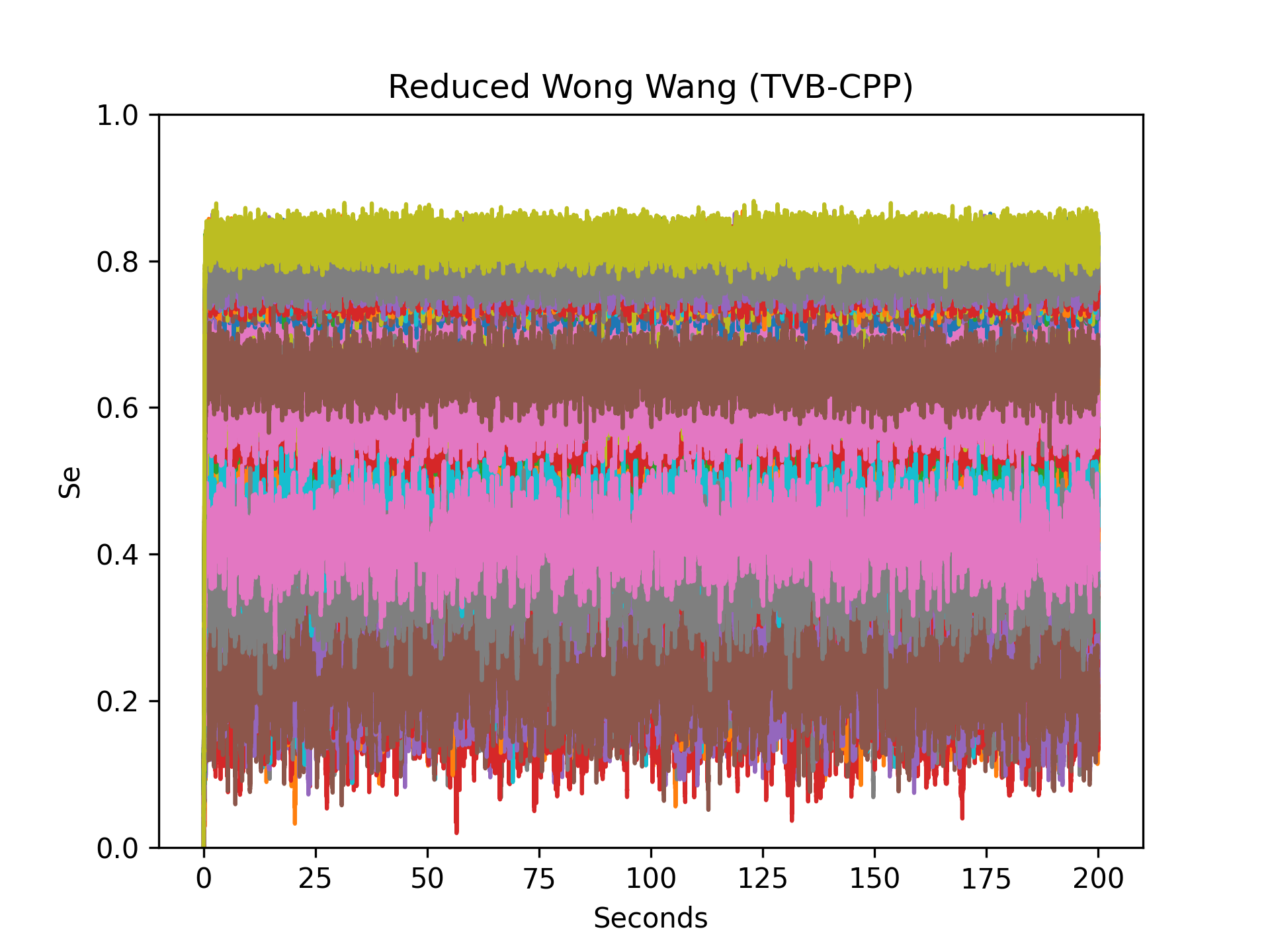}
\includegraphics[width=0.3\textwidth,trim={0.6cm 0 1cm 0},clip]{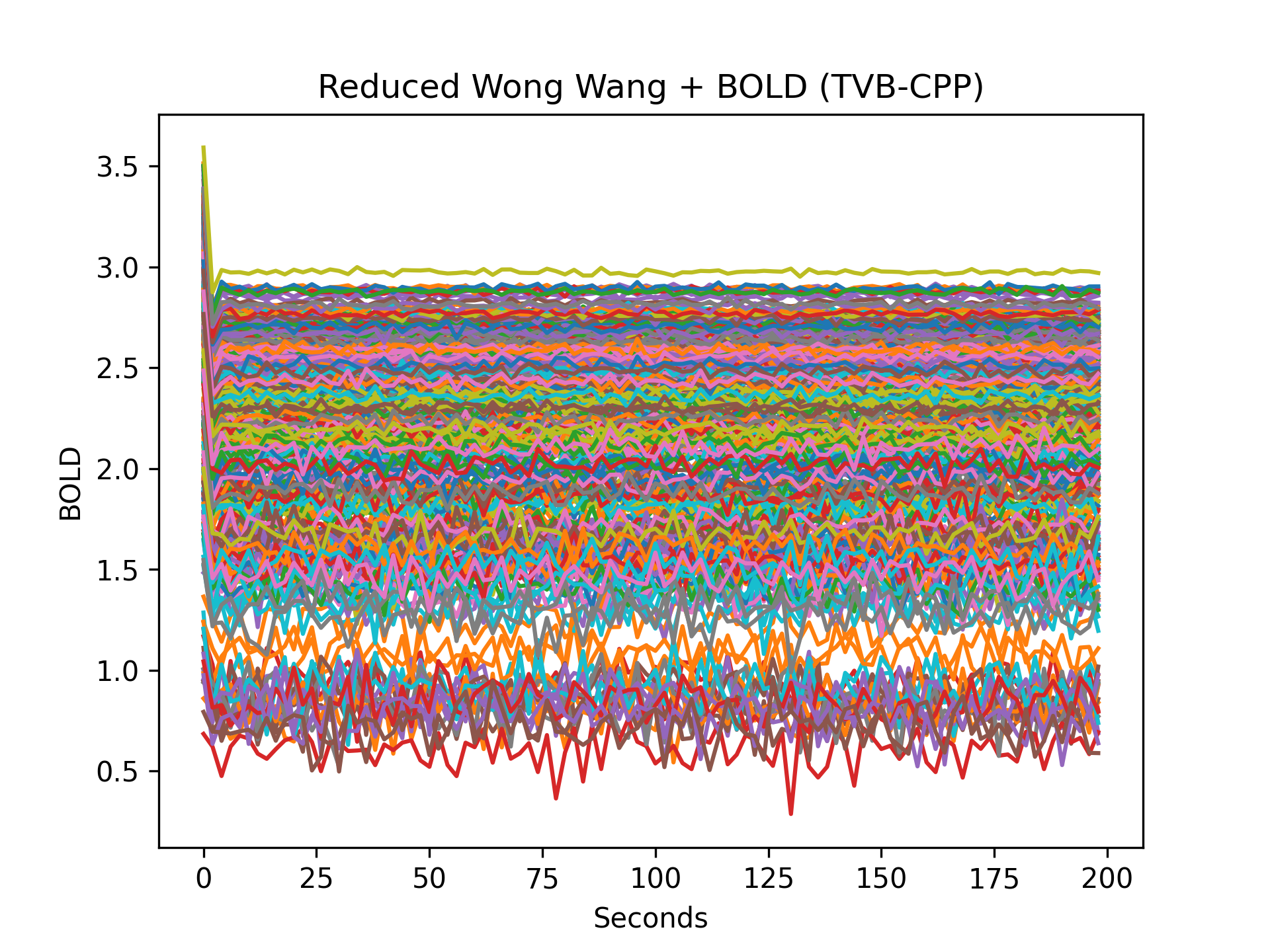}
\caption{Different simulations comparing how adding noise changes the results of the simulations with the Modified Balanced Excitation-Inhibition model.
Left column: 10s simulation comparing TVB and \fTVB{}.
Middle column: 200s simulation comparing TVB and \fTVB{}, both used to generate a BOLD signal. 
Right column: two examples of a BOLD signal computed from the previous simulation, using the same algorithm (standard TVB monitor convolution) with both TVB and \fTVB{}. Observe that in all cases, despite the random number generation is not exactly the same, all comparisons produce a very similar global result. Noise used is $\sigma = 1\mathrm{e}{-5}$ in both TVB and \fTVB{}.
}
\label{fig:plotNoise}
\end{figure}

As a first test, we compared \fTVB{} with TVB-Python version 2.8.1, with FAST\_TVB, and with RateML. In all cases, we use a reduced Wong Wang model~\cite{Deco2014} and the Julich-Brain probabilistic cytoarchitectonic maps as a microstructural reference parcellation from the multilevel Human Brain Atlas, of 379 regions. All simulations evaluate 10 seconds and use a deterministic integration with an integration delta of 0.1 ms. Execution times are shown in Table~\ref{table:simulationWW}. In Figure~\ref{fig:plotBold}, top row, we show the plot of the first state variables ($S_e$) of the 379-region model over 10 seconds. Both results shown are numerically identical. For the sake of completeness, although the random number generators in Python and C++ are different, we compared one of the models, the BEI model, with noise, both for TVB and \fTVB{}, leading to practically indistinguishable results except for the noise itself, see Figure~\ref{fig:plotNoise}.

\begin{table}[ht!]
\centering
\begin{tabular}{|c|c|c|} 
 \hline
 TVB & FAST\_TVB & \fTVB{} \\ [0.5ex] 
 \hline\hline
 40.1 & 5.1 & 6.4\\ [1ex] 
 \hline
\end{tabular}
\caption{Execution times for the reduced Wong Wang model simulation (in seconds).}
\label{table:simulationWW}
\end{table}

\paragraph{Feedback Inhibition Control:}
As a use case beyond brain simulations, we have implemented the Feedback Inhibition Control mechanism proposed by Deco et al.~\cite{Deco2014}. We have implemented the optimization method that allows finding the $J_i$ values that allow the neuron's firing rates to stay in a given interval, for every value of the global coupling value $G$. In particular, we have reproduced Figure 2C of Deco et al.~\cite{Deco2014}, see Figure~\ref{fig:plotBold}, lower right. In this case, the results are not numerically identical since we are not using the same data as used in the paper for the test, but, as can be seen, the outcome is completely equivalent.

\subsection{Mean Field Adaptive Integrate and Fire Neurons (mean-field AdEx)}

In this section, we will elaborate on the first and second-order Mean-Field Adaptive Integrate and Fire (mean-field AdEx) model.

\paragraph{First Order model:}
A mean-field model of conductance-based networks of adaptive exponential integrate-and-fire neurons~\cite{Zerlaut2018}
\begin{align*}
T \frac{\partial \nu_\mu}{\partial t} =& -F_\mu(\nu_e,\nu_i) + \nu_\mu,{\forall}\mu\in\{e,i\} \\
\frac{\partial W_k}{\partial t} =& W_k/tau_w - b*E_k 
\end{align*}
where $\nu_\mu=\{E,I\}$, and $F_\lambda$ is the Transfer Function, which should be tailored for each specific type of single neuron model desired. Carlu et al.~\cite{Carlu2020} used this framework to obtain Transfer Functions for diverse models such as nonlinear integrate-and-fire or Hodgkin-Huxley models.

\paragraph{Second Order model:}
A mean-field model of conductance-based networks of spiking neurons with adaptation~\cite{diVolo2019}. The general formulation for the \textit{\textbf{mean-field AdEx adaptation second order}} model (a.k.a., Zerlaut's model, di Volo's model, ...) as a dynamical unit at a node $k$ in a network with $l$ nodes reads $\forall \mu,\lambda,\eta \in \{e,i\}^3\,$:
\begin{align*}
T \, \frac{\partial \nu_\mu}{\partial t} = & (\mathcal{F}_\mu - \nu_\mu )
+ \frac{1}{2} \, c_{\lambda \eta} \,
\frac{\partial^2 \mathcal{F}_\mu}{\partial \nu_\lambda \partial \nu_\eta} \\
T \, \frac{\partial c_{\lambda \eta} }{\partial t}  =  & A_{\lambda \eta} +
(\mathcal{F}_\lambda - \nu_\lambda ) \, (\mathcal{F}_\eta - \nu_\eta ) + \\
& c_{\lambda \mu} \frac{\partial \mathcal{F}_\mu}{\partial \nu_\lambda} +
c_{\mu \eta} \frac{\partial \mathcal{F}_\mu}{\partial \nu_\eta}
- 2  c_{\lambda \eta}\\
\frac{\partial W_k}{\partial t} = & W_k/tau_w-b*E_k
\end{align*}
with:
\begin{equation*}
   A_{\lambda \eta} =
        \begin{cases}
        \frac{\mathcal{F}_\lambda \, (1/T - \mathcal{F}_\lambda)}{N_\lambda}
        \qquad & \textrm{if  } \lambda=\eta \\
        0 \qquad & \textrm{otherwise}
        \end{cases} 
\end{equation*}

This model has recently been extended to Whole-Brain modeling~\cite{Goldman2023} by adding the coupling through the structural connectivity into the synaptic current.

\begin{figure}[!ht]
\centering
    \begin{subfigure}[b]{1.\linewidth}
        \centering
        \includegraphics[width=0.49\textwidth,trim={0.7cm 0 1cm 0},clip]{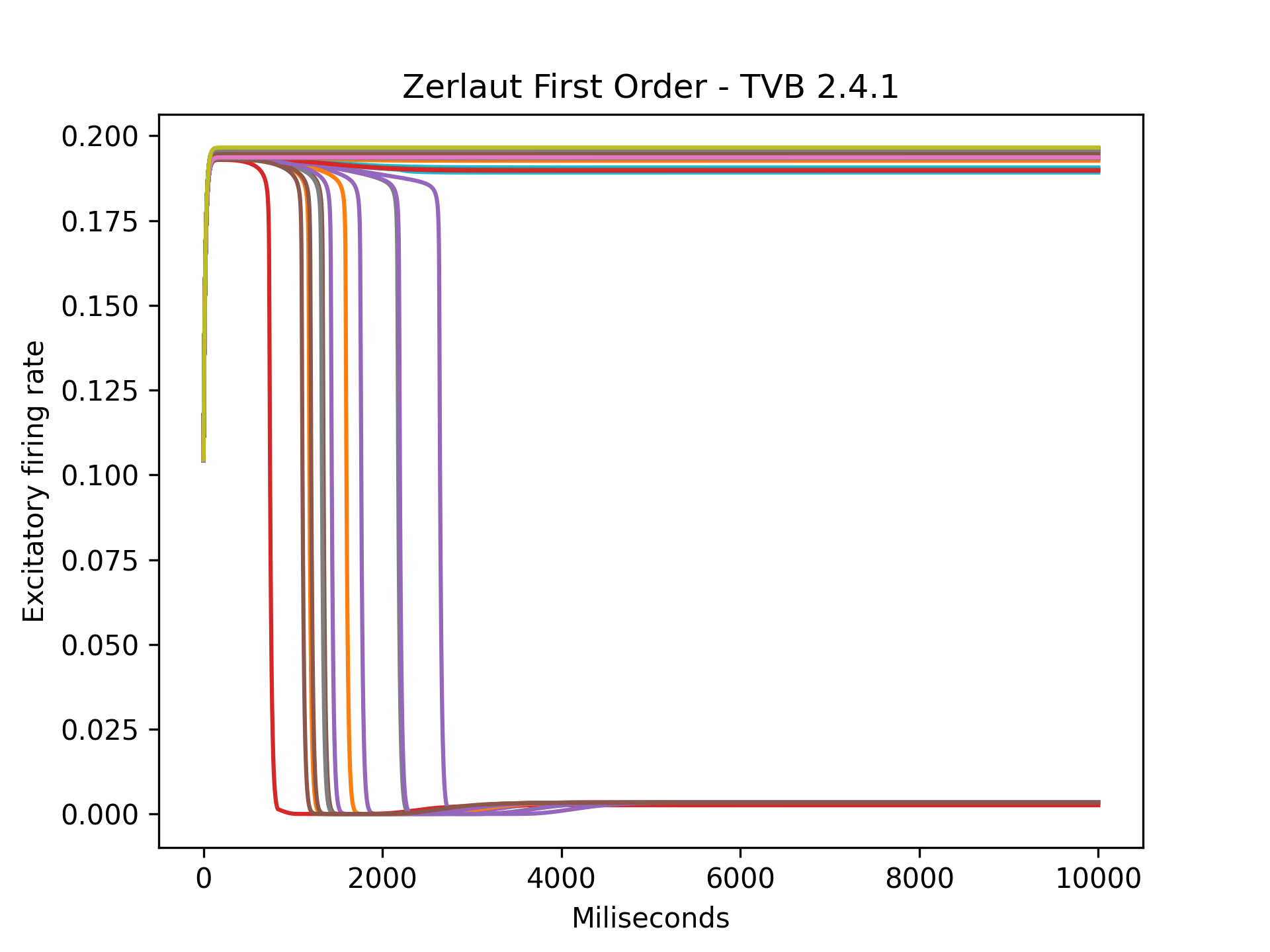}
        \includegraphics[width=0.49\textwidth,trim={0.7cm 0 1cm 0},clip]{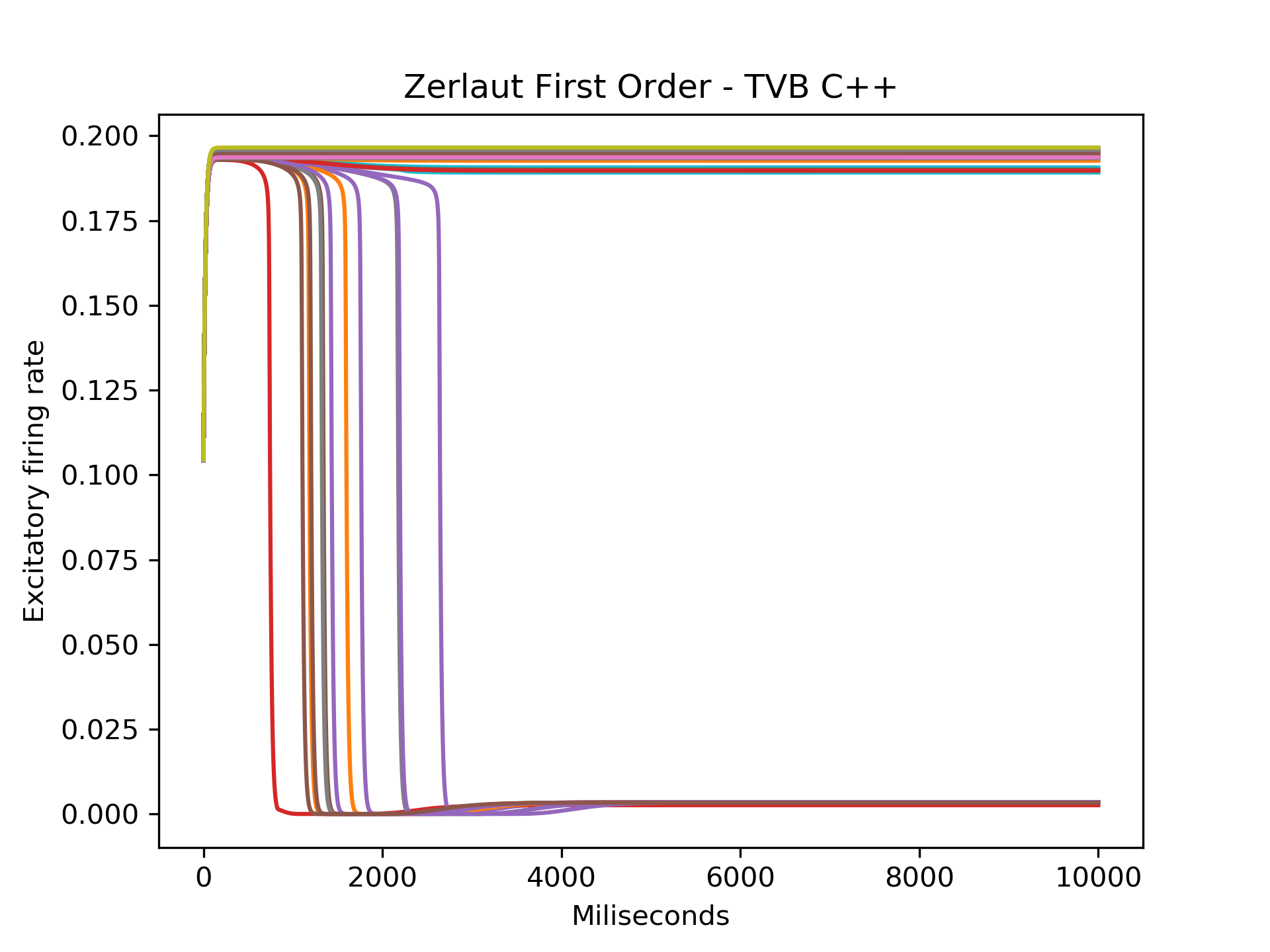}
        \caption{}\label{fig:plotZ}
    \end{subfigure}
    \begin{subfigure}[b]{0.49\linewidth}
        \centering
        \includegraphics[width=1.\linewidth,trim={0.8cm 0 1cm 0},clip]{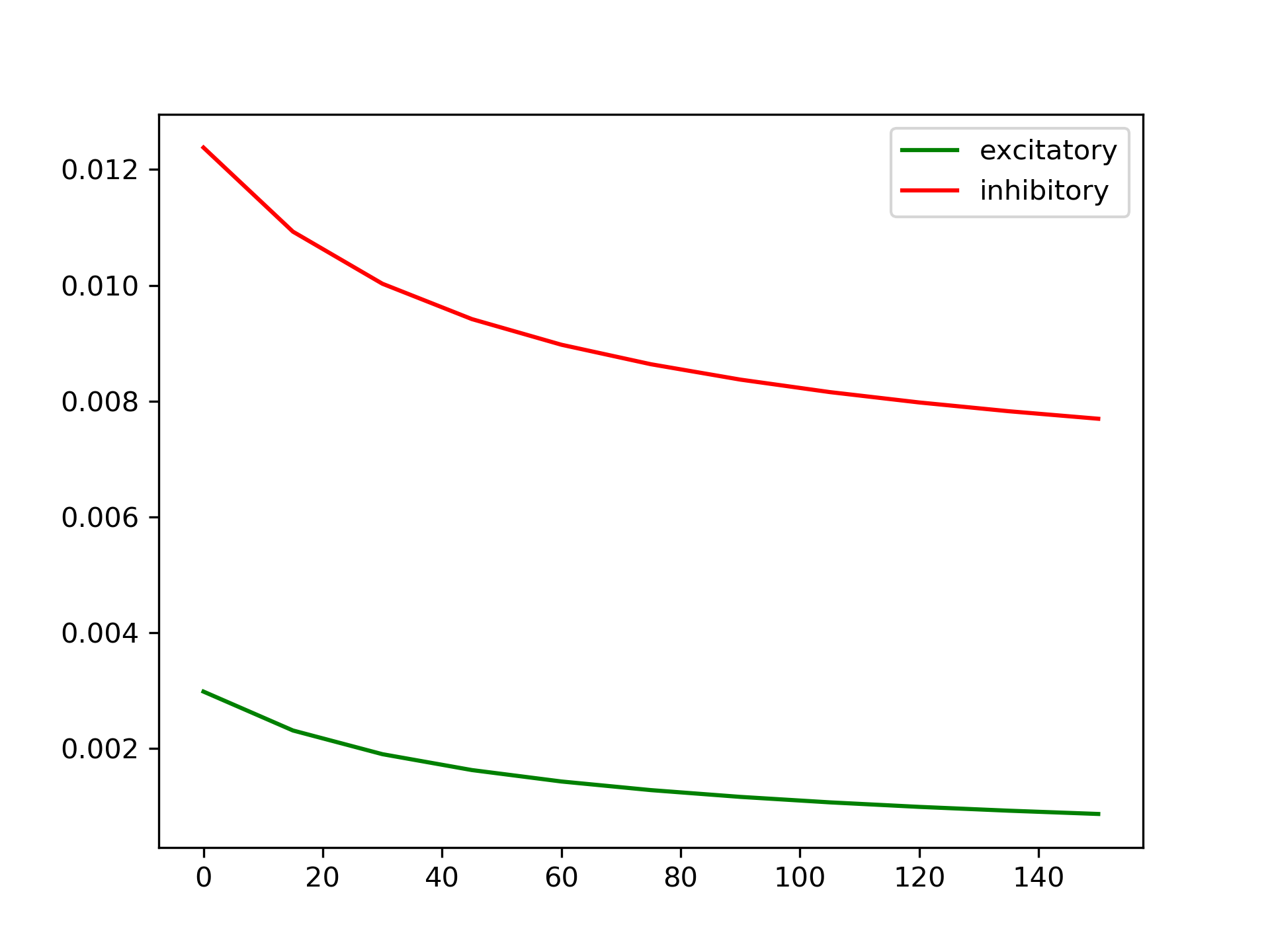}
        \caption{}\label{fig:zerlaut_fig2a}
    \end{subfigure}
    \begin{subfigure}[b]{0.49\linewidth}
        \centering
    \includegraphics[width=1.\linewidth,trim={0.8cm 0 1cm 0},clip]{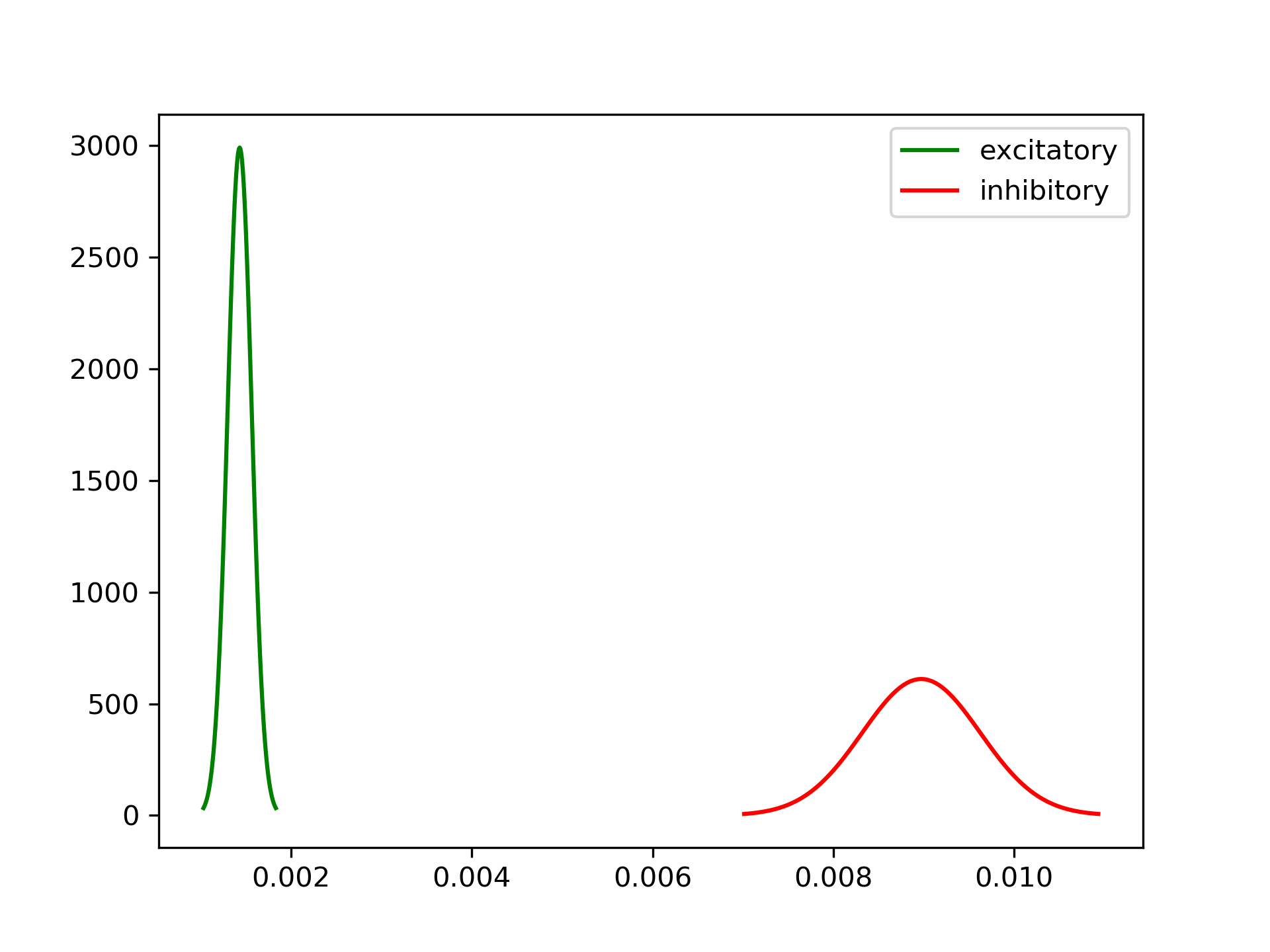}
    \caption{}\label{fig:zerlaut_fig2b}
    \end{subfigure}
\caption{Reproducing the results of mean-field AdEx module from di Volo et al.~\cite{diVolo2019}. 
(a) Plot of a whole brain simulation with TVB and \fTVB{} for the mean-field AdEx first-order model.
(b) the stable firing rate of the model for different values of the adaptation parameter $b_e$. 
(c) the firing rate distribution is predicted by the model for $b_e=60$.} 
\label{fig:zerlaut_fig2}
\end{figure}

We have also performed tests with other, more complex models, like the mean-field AdEx second order model~\cite{Zerlaut2018,diVolo2019}. In this case, comparison can only be done with TVB, since FAST\_TVB only works with the reduced Wong Wang model and does not allow new models to be easily incorporated. In both cases (i.e., TVB and \fTVB{}), testing another model involves only the change of a single line of code. In Figure~\ref{fig:plotZ} we show the excitatory firing rate of the 379-region model over 10 seconds. Again, both results are numerically identical.

\begin{table}[!ht]
\centering
\begin{tabular}{|c|c|} 
 \hline
 TVB & \fTVB{} \\ [0.5ex] 
 \hline\hline
 489.65 & 23.5 \\ [1ex] 
 \hline
\end{tabular}
\caption{Execution times for the mean-field AdEx first order model simulation (in seconds).}
\label{table:simulationZ}
\end{table}

Execution times are shown in Table~\ref{table:simulationZ}. As can be seen, simulation times are speeded up about 10x times with \fTVB{}. This takes into account that TVB Python uses Numba to accelerate the core of the simulation and the computation of the derivatives of the differential equations of the model. In \fTVB{} we have not applied any further specific optimization.

Thus, the last set of tests has been performed using the mean-field AdEx model~\cite{Boustani2019,Zerlaut2018,diVolo2019}. In this case, we have reproduced some of the results presented in~\cite{diVolo2019}, specifically Figure 2 in the paper. The results are shown in Figure~\ref{fig:zerlaut_fig2}.

\subsection{FRE Model}
\label{sec:montbrio}

Montbrió and coworkers~\cite{Montbri2015,Montbrio_2020}, proposed a method to derive the Firing Rate Equations (FREs) for networks of heterogeneous, all-to-all coupled quadratic integrate-and-fire (QIF) neurons, which is exact in the thermodynamic limit, i.e., for large numbers of neurons. Their formulation uses mean-field equations to reveal a dual role of electrical synapses: First, they equalize membrane potentials favoring the emergence of synchrony. Second, electrical synapses act as ``virtual chemical synapses'', which can be either excitatory or inhibitory depending upon the spike shape. The original results offer a precise mathematical explanation of the intricate effect of electrical synapses in collective synchronization. 
\begin{equation}
    \begin{gathered}
        \tau \frac{dr}{dt} = \frac{\Delta}{\tau\pi} + 2 r u - g r \\
        \tau \frac{du}{dt} = u^2 + \bar{\eta} - (\pi\tau r)^2 + (J+g \ln a) \tau r 
    \end{gathered}
    \label{eq:FRE}
\end{equation}
The parameters $\bar{\eta}$ and $\Delta$ are, respectively, the center and width of the static distribution of inputs to the individual spiking neurons, which is considered Lorentzian. Also, we have the recurrent inputs with their synaptic weight $J$, and the electric synapses weighted by their respective weight $g$.

This model can be extended to an Excitatory-Inhibitory (E-I) network of quadratic integrate-and-fire neurons modeling a Pyramidal-Interneuronal Network Gamma (PING) rhythm~\cite{Dumont2019,Reyner-Parra2022}, and then extend it again to a whole-brain made of these E-I nodes, by formulating the system of equations:
\begin{equation}
    \begin{gathered}
        \tau^e \frac{dr_j^e}{dt} = \frac{\Delta^e}{\tau^e\pi} 
                                 + 2 r_j^e u_j^e - g^e r_j^e \\
        \tau^i \frac{dr_j^i}{dt} = \frac{\Delta^i}{\tau^i\pi} 
                                 + 2 r_j^i u_j^i - g^i r_j^i \\
        \tau^e \frac{du_j^e}{dt} = {u_j^e}^2 + \bar{\eta^e} 
                                 - (\pi\tau^e r_j^e)^2 + I_j^e \\
        \tau^i \frac{du_j^i}{dt} = {u_j^i}^2 + \bar{\eta^i} 
                                 - (\pi\tau^i r_j^i)^2 + I_j^i \\
        \tau^N \frac{dS_j^{ee}}{dt} = - S_j^{ee} + J^{N(ee)} r_j^e \\
        \tau^N \frac{dS_j^{ie}}{dt} = - S_j^{ie} + J^{N(ie)} r_j^e \\
        I_j^e = I_{ext}^e + \tau^e S_j^{ee} 
            - J_j J^{G(ei)} \tau^i r_j^i  
            + G J^A \tau^e \sum_{k\neq j} C_{kj} S_k^{ee} \\
        I_j^i = I_{ext}^i + \tau^e S_j^{ie} - J^{G(ii)} \tau^i r_j^i
    \end{gathered}
    \label{eq:AMPA-GABA-NMDA}
\end{equation}

Observe that, in this last expression, we have the long-range synaptic weight $J^A$, identified with the AMPA current. Also, we  introduced the slower NMDA gating variable $S$, which modulates connections from the excitatory population into both the excitatory and inhibitory populations, being regulated by its respective time constant $\tau^N$. Finally, the inhibitory connections have been associated with GABA, so we have labeled its constant as $J^G(xi)$, with $x$ being the receiving population (excitatory $e$ or inhibitory $i$). From this system, it is easy to recover the equations used by Sanz Perl and co-workers~\cite{SanzPerl2023} by removing the electrical coupling ($g^e=g^i=0$), removing the FIC mechanism, assuming no external stimulation ($I_{ext}^e = I_{ext}^i$ = 0), considering that $\tau^e = \tau^i = \tau^m$, equating $J^{long,e} = J^{ee}$, and considering the steady state of all the gating variables $S^{e,i}$.

\begin{figure}[ht!]
    \centering
    \begin{subfigure}[b]{1.\linewidth}
        \centering
            \includegraphics[width=.3\textwidth]{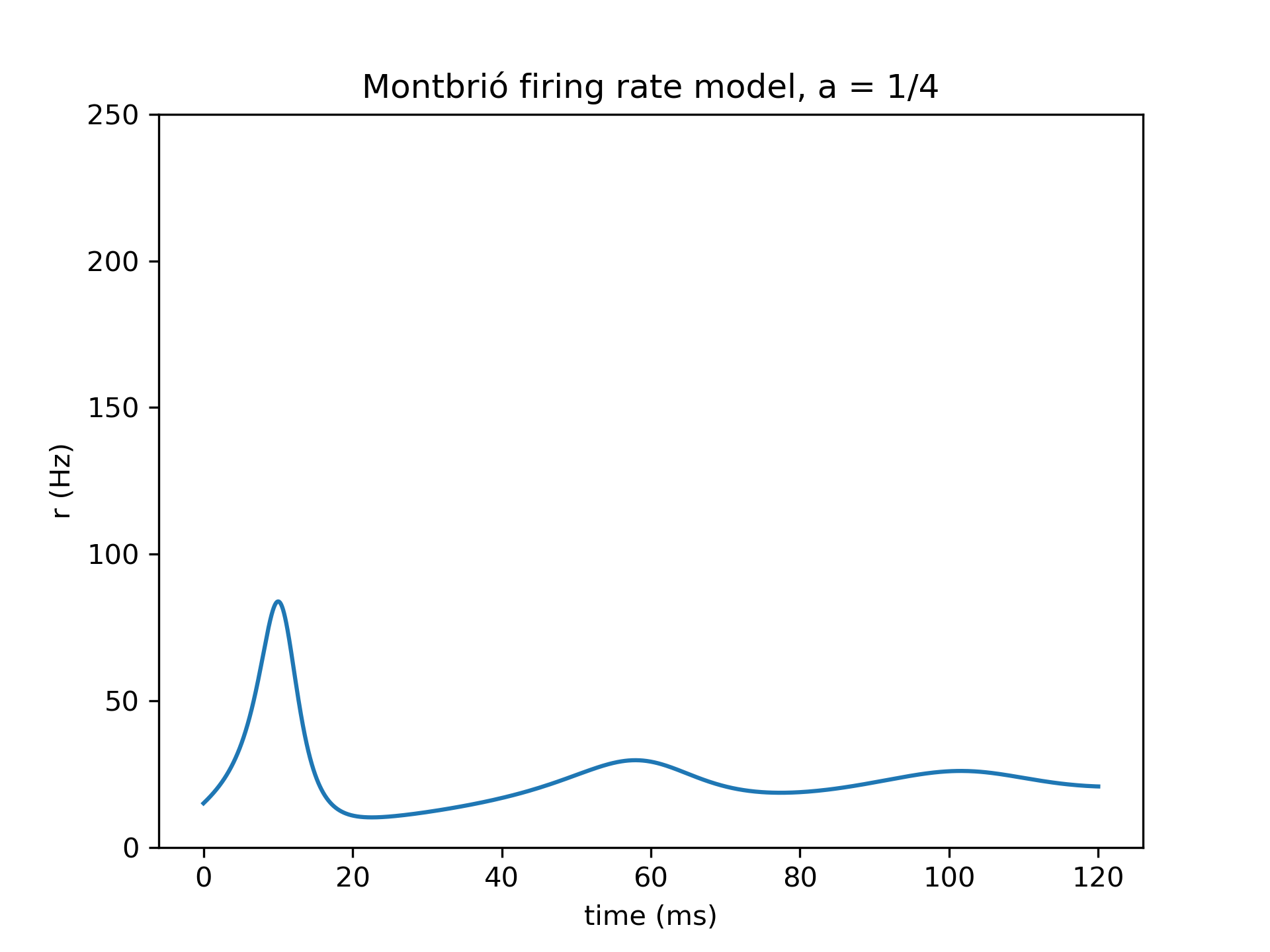}
            \includegraphics[width=.3\textwidth]{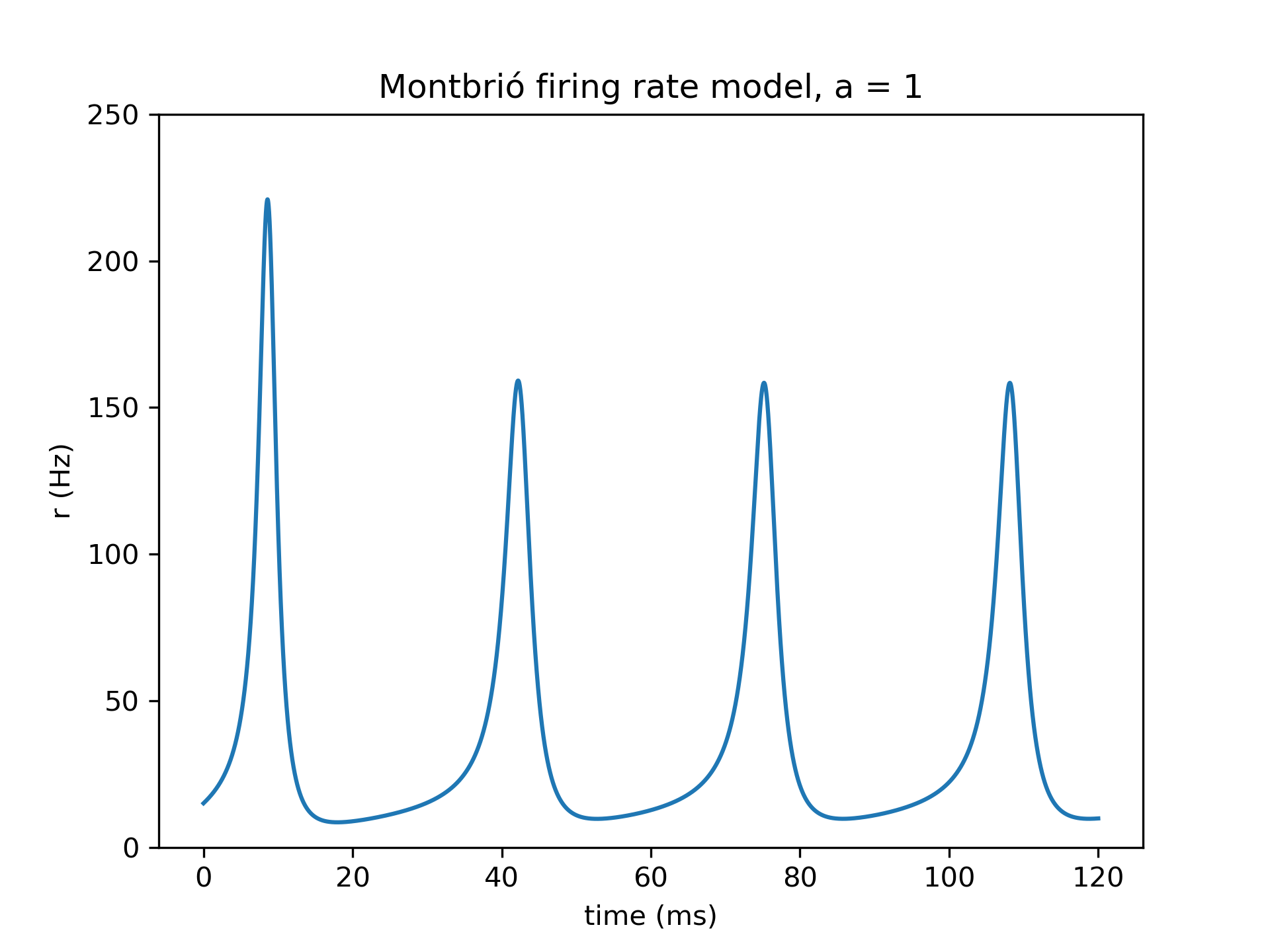}
            \includegraphics[width=.3\textwidth]{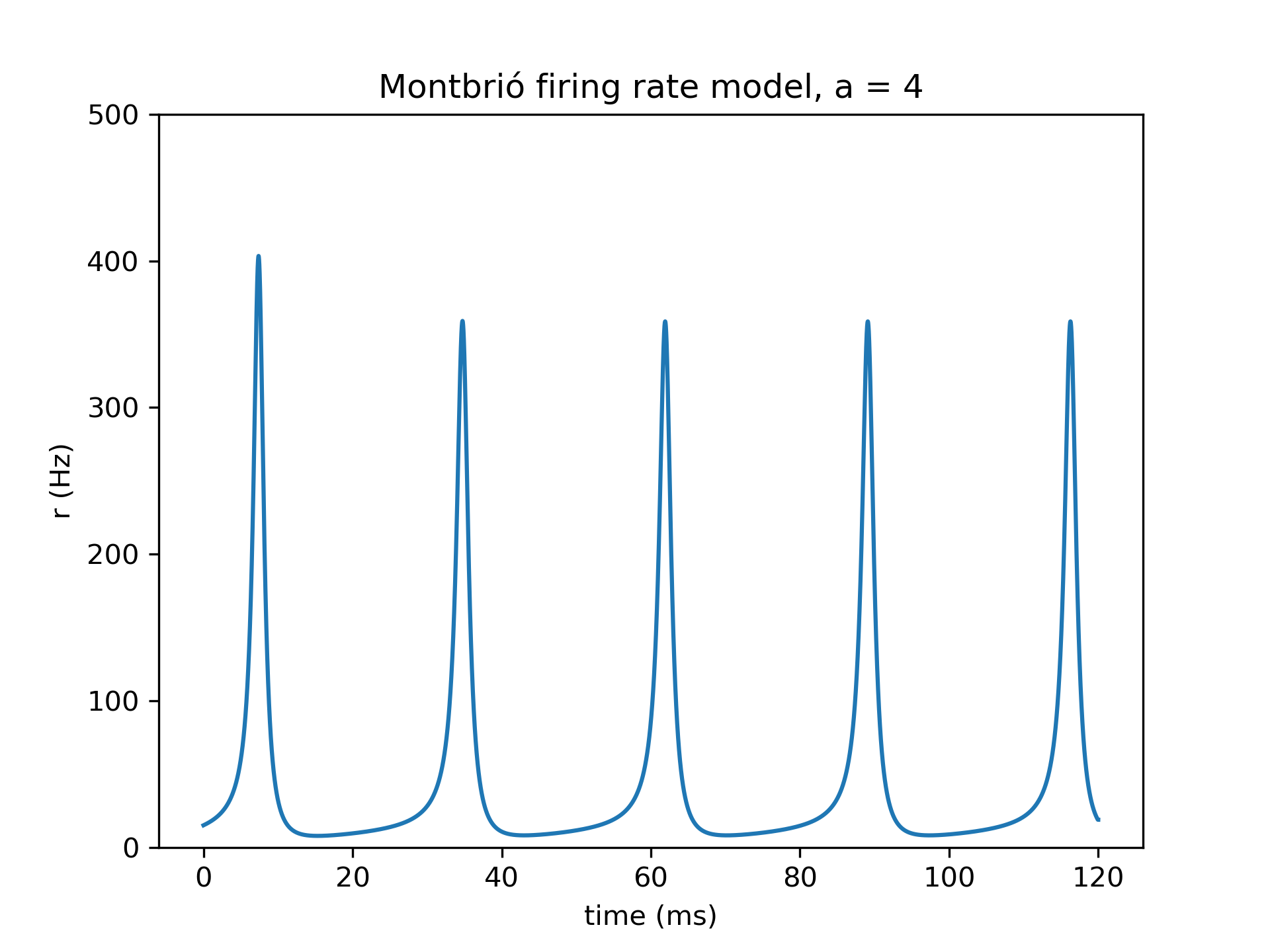}
        \caption{}\label{fig:montbrio_fig3}
    \end{subfigure}
    \begin{subfigure}[b]{0.49\linewidth}
        \centering
        \includegraphics[width=1.\linewidth]{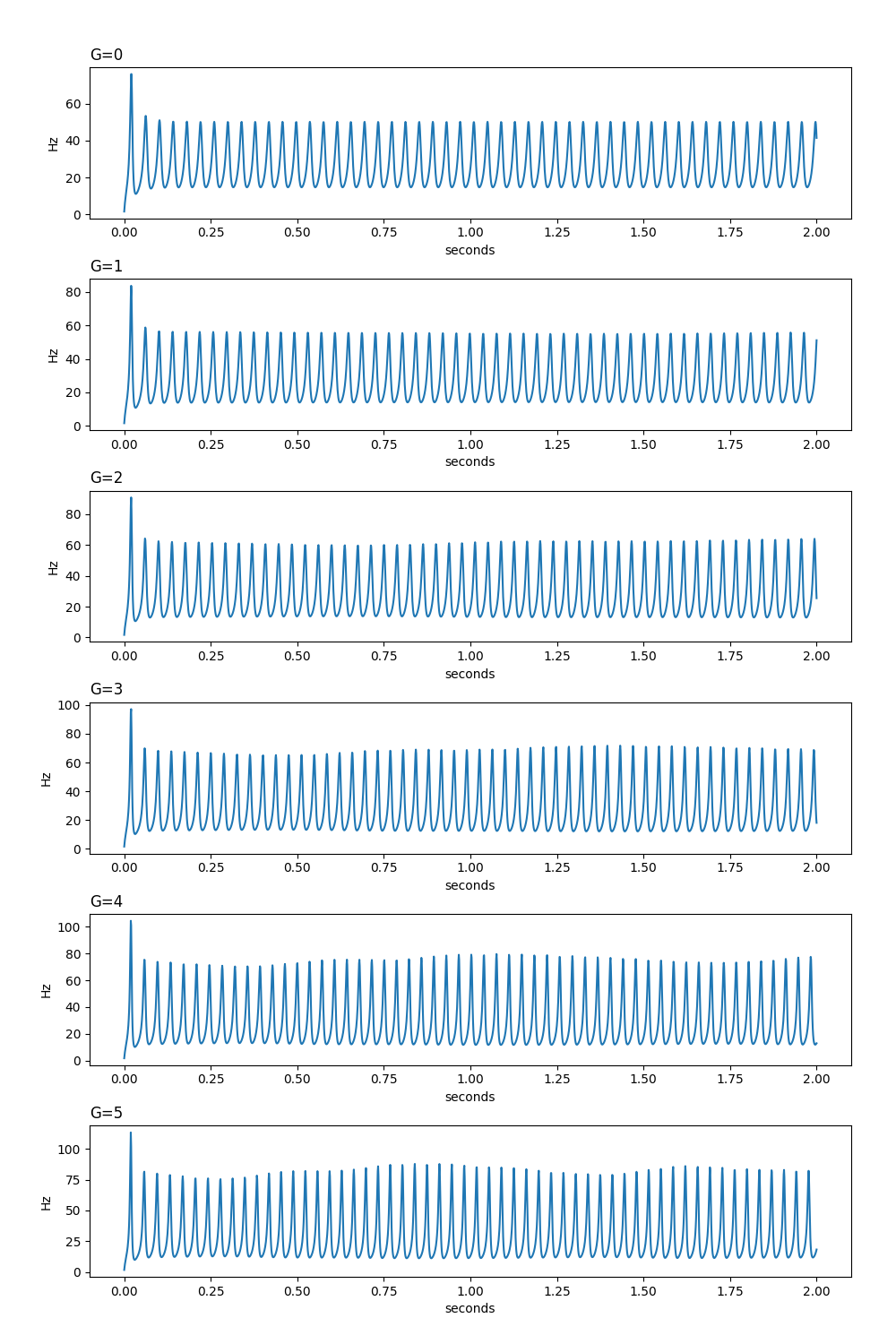}
        \caption{}\label{fig:montbrio_fig3a}
    \end{subfigure}
    \begin{subfigure}[b]{0.49\linewidth}
        \centering
    \includegraphics[width=1.\linewidth]{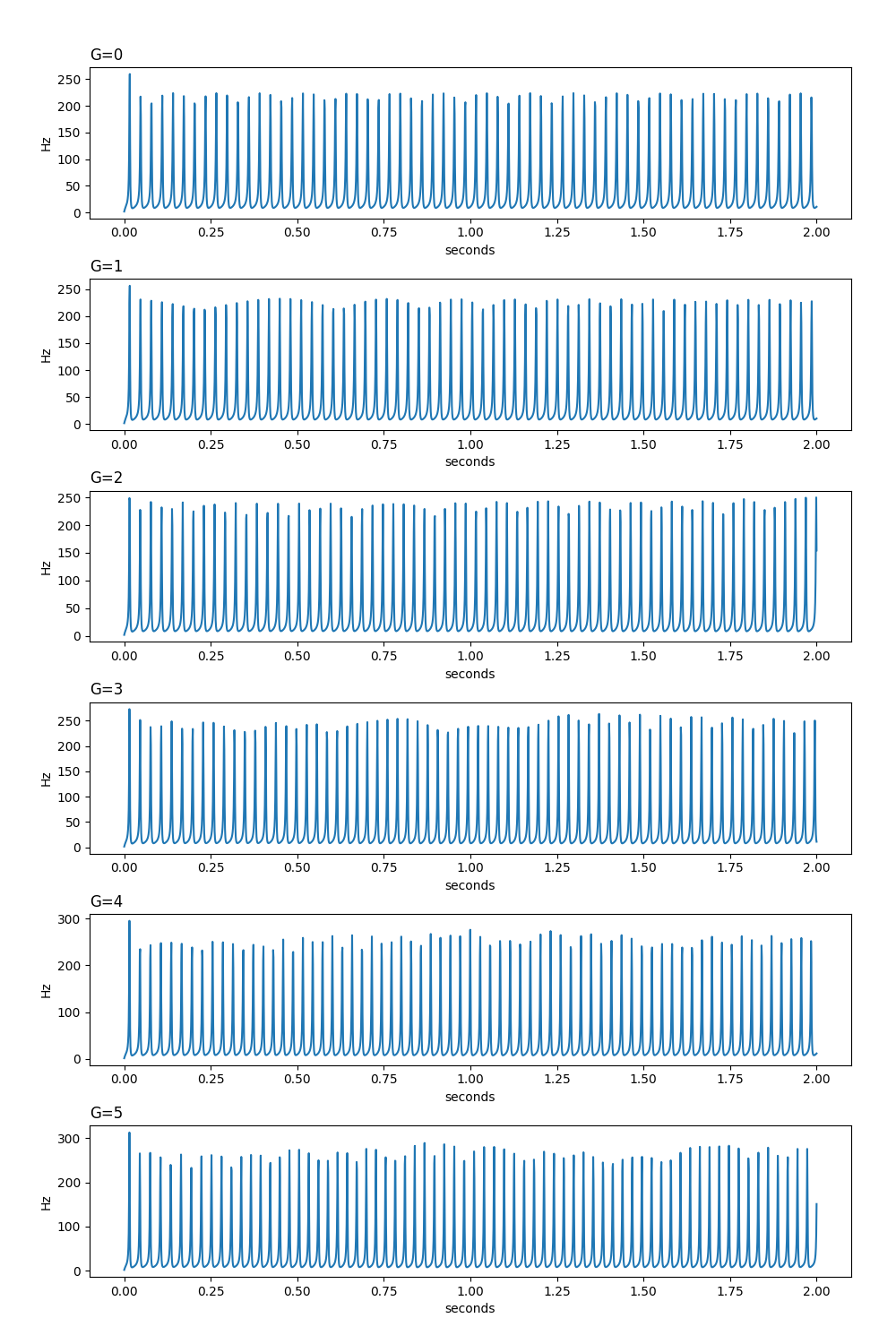}
    \caption{}\label{fig:montbrio_fig3b}
    \end{subfigure}
\caption{Using \fTVB to reproduce the result of a whole brain simulation using the FRE model~\cite{Montbrio_2020} (Figure 3 in the manuscript). 
(a) Three values of the spike asymmetry parameter $a$ are used: $1/4$, $1$ and $4$.
(b-c) Using a Desikan parcellation with 70 ROI, setting the spike asymmetry parameter $a$ at $1/4$ (b) and $1$ (c) and showing the variation in the firing rate when the coupling parameter $G$ varies from $0$ to $5$.
}
\label{fig:montbrio_fig3_beyond}
\end{figure}

Once the full whole-brain model is implemented, to recover the original formulation by Montbri\'o and coworkers~\cite{Montbrio_2020}, we need, first, to decouple all populations by setting $G=0$, then decouple the inhibitory and excitatory populations by choosing $J^G=0$, and finally approximate the behavior of the gating variable $S$ by its steady state behavior. However, this last step requires setting $\frac{dS}{dt}=0$, which is not possible without modifying the code. However, depending on the integration algorithm used, we could set $\tau^N = \Delta t$, which, for Euler-Maruyama will reduce to the correct expression for $S$, and for other more advanced integrators (e.g., Runge-Kutta) will lead to an average of the firing rate during the current time-step. In particular, we have reproduced Figure 3 from the original paper, using three values of the spike asymmetry parameter $a$: $0.25$, $1$, and $4$ (see Figure \ref{fig:montbrio_fig3}).

Using the same settings, we can go beyond the results presented in the original papers and extend the model to a whole-brain simulation. The results for the behavior of one node, when coupled with all the others for values of $G$ of $1/4$ and $1$, are shown in Figure~\ref{fig:montbrio_fig3_beyond}. As expected, the behavior for a given node changes significantly when coupled with the rest of the network, showing significant oscillatory behavior.

\subsection{BOLD simulations}
Another basic feature of \fTVB{} is the ability to use a BOLD monitor to compute fMRI signals from the raw output of the models, such as the models presented before. FAST\_TVB also computes a BOLD signal from the reduced Wong Wang model but is limited to the Balloon-Windkessel hemodynamic model. We have compared BOLD signal generation in our model with both TVB 2.8.1 and FAST\_TVB. As already mentioned, since the details of the BOLD algorithm used are different for both of these systems (see above), \fTVB{} implements both of them to perform fair comparisons.

\subsubsection{BOLD monitor using HRF convolutions}
TVB 2.8.1 implements, as its main option, a hemodynamic response function (HRF) to compute the BOLD signal. In the current implementation, these HRF kernels are implemented using a first-order Volterra kernel. To test BOLD generation we have used a smaller parcellation to have shorter execution times. In particular, we used the Desikan parcellation with 68 regions. Execution times are shown in Table~\ref{table:boldTVB}. In figure~\ref{fig:plotBold}, middle row, we show the plot of a BOLD signal over 2 minutes. As can be seen, both are numerically identical. 

\begin{table}[!ht]
\centering
\begin{tabular}{|c|c|} 
 \hline
 TVB & \fTVB{} \\ [0.5ex] 
 \hline\hline
 250.895 & 4.41 \\ [1ex] 
 \hline
\end{tabular}
\caption{Execution times for the reduced Wang Wong BOLD generation, using the Desikan parcellation (68 ROIs) and with the HRF convolution-based model.}
\label{table:boldTVB}
\end{table}

From Table~\ref{table:boldTVB} we can observe a speedup of about 50x, which includes both the Reduced Wong Wang model simulations, \emph{and} the BOLD simulation. This result is mainly because the BOLD computations in TVB do not use any kind of acceleration (e.g., using the \emph{numba} library) and are pure Python. In these cases, the performance difference can be quite dramatic.

\subsubsection{BOLD monitor using Balloon-Windkessel}
We have also compared BOLD generation from electrical activity obtained from a simulation with FAST\_TVB. In this case, both FAST\_TVB and \fTVB{} use a Balloon-Windkessel algorithm. We have used the same smaller parcellation to have a meaningful comparison. Execution times are shown in Table \ref{table:boldFAST}. In Figure~\ref{fig:plotBold}, first row, we show the plot of the BOLD signal over 2 minutes. Again, both signals are numerically identical. 

\begin{table}[!ht]
\centering
\begin{tabular}{|c|c|} 
 \hline
 FAST\_TVB & \fTVB{} \\ [0.5ex] 
 \hline\hline
 10.89 & 4.26 \\ [1ex] 
 \hline
\end{tabular}
\caption{Execution times for the reduced Wang Wong BOLD generation using the Balloon-Windkesssel model, with the Desikan parcellation (68 ROI).}
\label{table:boldFAST}
\end{table}

The 3x speedup found here could be explained because, with this parcellation and this model, the BOLD computations take the bulk of the total time, and these operations are vector and matrix intensive, and we strongly believe the optimizations provided by the Eigen library are responsible.

Up to this point, all comparisons have been conducted within a single-threaded process. However, several algorithms necessitate a fitting process of model parameters, thereby requiring parallel execution of numerous simulations. In this regard, both FAST\_TVB and \fTVB{} are equipped to leverage multi-core CPUs for parallel execution. Nevertheless, for extensive parallel executions, the utilization of GPUs proves to be more efficient.

\begin{figure}[ht!]
    \centering
    \begin{subfigure}[b]{0.49\linewidth}
        \centering
        \includegraphics[width=1.\linewidth]{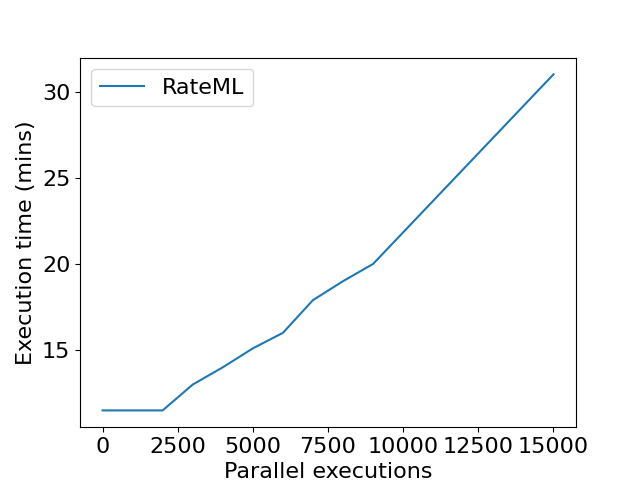}
        \caption{}\label{fig:ratemv_cpp_figa}
    \end{subfigure}
    \begin{subfigure}[b]{0.49\linewidth}
        \centering
    \includegraphics[width=1.\linewidth]{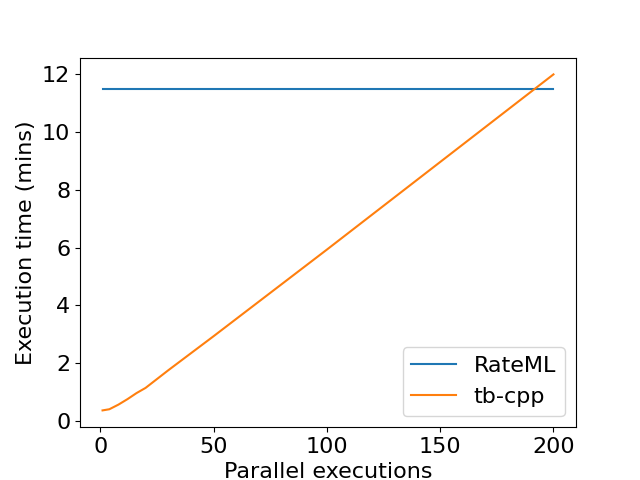}
    \caption{}\label{fig:ratemv_cpp_figb}
    \end{subfigure}
\caption{On the left panel we see the execution times for RateML as the number of parallel executions grows. From 1 to 2000 parallel runs, the time is almost constant around 11.5 minutes. However, when the GPU limit is reached, the time starts to behave linearly with the number of executions. On the right panel, the execution times of RateML against \fTVB{}. We can observe a linear behavior for \fTVB{} almost since the beginning, being more efficient up to 180/190 parallel executions, when RateML outperforms it given the massively parallel characteristic of modern GPUs.} 
\label{fig:ratemv_cpp_fig}
\end{figure}

The TVB project has introduced a GPU backend named RateML, which entails rewriting the model derivative equations to enable their use within a CUDA kernel. A comparison between RateML and \fTVB{} is presented in Figure~\ref{fig:ratemv_cpp_fig}. The findings demonstrate that for a small number of parallel executions (up to 180/190), \fTVB{} exhibits superior performance, especially at lower numbers. However, in scenarios requiring massive parameter sweeps, RateML surpasses \fTVB{} by an order of magnitude in terms of efficiency.

\section{Summary and Discussion}\label{sec:Conclusions}
As demonstrated, \fTVB{} surpasses the original TVB Python implementation in terms of speed and performance while preserving its flexibility. The primary role of \fTVB{} as an efficient Back-End for demanding computational tasks is seamless and economical, thanks to its Python bindings tailored for ease of use and adaptability.
It is noteworthy that \fTVB{} holds significant potential for further optimization, leveraging available libraries such as Intel MKL or CUDA. Additionally, unlike Python, which is hindered by the Global Interpreter Lock, the C++ version of \fTVB{} can effortlessly exploit native threads. The current iteration of \fTVB{} offers a straightforward thread pool mechanism to capitalize on multicore architectures.
\fTVB{} constitutes a valuable addition to the TVB family owing to its high compatibility, exceptional flexibility, and superior performance. It emerges as an ideal backend for computationally intensive tasks, requiring no additional expertise beyond TVB basics.

In conclusion, \fTVB{} stands as a formidable asset in the realm of computational neuroscience, offering a compelling blend of speed, flexibility, and ease of use. Through meticulous optimization and seamless integration with the renowned TVB platform, \fTVB{} elevates the capacity for large-scale simulations and heavy-duty computational tasks to unprecedented heights.
Its exceptional performance, outpacing the original TVB-Python implementation, underscores its prowess as a backend solution for tackling complex neuroscientific inquiries with efficiency and precision. Moreover, \fTVB{}'s compatibility and intuitive Python bindings ensure a smooth transition for users, requiring no additional expertise beyond fundamental TVB knowledge.

With the potential for further optimization and the ability to harness advanced computing resources like supercomputers, \fTVB{} emerges as an indispensable tool for researchers seeking to push the boundaries of computational neuroscience. By seamlessly marrying speed, flexibility, and ease of use, \fTVB{} paves the way for groundbreaking discoveries and advancements in our understanding of the brain's intricate dynamics.

\section*{Code availability}
All the code described in this paper can be found in a Github public repository~\cite{Martin_TVB_C_Library_2022}.

\section*{Author contributions (CRediT)}
PR, VJ and GD: conceptualization, methodology. 
IM, GZ, JF, MS and GP: methodology. 
IM: development, software.
IM, GP, GZ, GD, MS, PR: Visualization, Investigation.
PR, VJ, GP: Formal Analysis.
IM, GZ, GP: Writing – original draft.
GP: Funding acquisition.
All authors: Validation, Data curation, Writing – review and editing

\section*{Acknowledgements}
We thank Michiel van der Vlag for assisting in the comparisons to Rate ML. We thank Alain Destexhe, Damien Depannemaecker and Ernest Montbri\'o for their help adapting the Mean-field AdEx and the ERP population models.

\section*{Funding}
IM and GP were partially funded by Grant PID2021-122136OB-C22 funded by MCIN/AEI/10.13039/501100011033 and by ERDF A way of making Europe.
This work has been supported by the European Union’s Horizon 2020 research and innovation program under Specific Grant Agreement No. 785907 (HBP SGA2) and Specific Grant Agreement No. 945539 (Human Brain Project SGA3).
This work was supported by the Virtual Research Environment at the Charité Berlin – a node of EBRAINS Health Data Cloud. Part of computation has been performed on the HPC for Research cluster of the Berlin Institute of Health. PR acknowledges support by EU Horizon Europe program Horizon EBRAINS2.0 (101147319), Virtual Brain Twin (101137289), EBRAINS-PREP 101079717, AISN – 101057655, EBRAIN-Health 101058516, Digital Europe TEF-Health 101100700; German Research Foundation SFB 1436 (project ID 425899996); SFB 1315 (project ID 327654276); SFB 936 (project ID 178316478; SFB-TRR 295 (project ID 424778381); SPP Computational Connectomics RI 2073/6-1, RI 2073/10-2,RI 2073/9-1; DFG Clinical Research Group BECAUSE-Y 504745852, PHRASE Horizon EIC grant; 101058240; Berlin Institute of Health \& Foundation Charité,

\section*{Conflict of Interest Statement}
The authors declare that the research was conducted in the absence of any commercial or financial relationships that could be construed as a potential conflict of interest.

\bibliography{refs}

\newpage
\onecolumn 
\section*{Figure Labels}

\clearpage
\begin{lstlisting}[language=Python, caption={Python bindings example. The output from this code gives the results shown in Figure~\ref{fig:montbrio_fig3_beyond}.}, label={code:PythonBindings}]
import tvbcpp
import numpy as np
import matplotlib.pyplot as plt
import scipy.io as sio

dt = 0.1 # Integration step in [ms]
m = sio.loadmat("CNT_S01_structure.mat")["CNT_S01_structure"]
m = m / np.max(m) * 0.2 # normallize matrix

# Set the structural connectivity matrix
tvbcpp.set_weights(m)
n_rois = m.shape[0]

print(f"Loaded connectivity matrix with {n_rois} number of ROIs")

tvbcpp.set_model("Montbrio")  # Choose the neuronal model

# Configure the integrator.
sigmas = np.array([0.0, 0, 0, 0, 1e-3, 1e-3])
tvbcpp.set_integrator("EulerStochastic", dt, sigmas)

# Configure the initial state of the model
n_state_vars = 6
initial_state = np.zeros((n_rois, n_state_vars))
tvbcpp.set_initial_state(initial_state)

tvbcpp.set_global_coupling(80.1)  # Set the global coupling

tvbcpp.set_model_parameter("delta_i", 1.18)  # Homogeneous working point [SanzPerl'23]

T = 300000 # simulation time in [ms]
tvbcpp.add_temporal_average_monitor(1.0, [0])  # temporal average monitor

result = tvbcpp.run_sim(0, T)  # Run the simulation

data_raw = result[0] # get the result of the first monitor
time_series = data_raw[1][:, :, 0]

tr = 2000 # BOLD sampling rate in [ms]

bold_2007b = tvbcpp.create_bold("Stephan2007b", tr)  # Create a BOLD object.
# Compute the BOLD signal from a time series of raw data
ts_2007b, data_bold_2007b = bold_2007b.compute_bold(time_series, 1.0)

# Subsample originial time series from Stephan2007b
n_2007b = T // tr
t_sub_2007b = np.zeros((n_2007b,))
data_sub_2007b = np.zeros((n_2007b, data_bold_2007b.shape[1]))
for n in range(n_2007b):
    t_sub_2007b[n] = np.mean(ts_2007b[n*tr:(n+1)*tr])
    data_sub_2007b[n, :] = np.mean(data_bold_2007b[n*tr:(n+1)*tr, :], axis=0)

\end{lstlisting}

\newpage
\begin{lstlisting}[language=Python, label={code:PythonPlotResults}, caption={Code to plot the output from Listing~\ref{code:PythonBindings}}]
# Plot for run_sim
fig, axs = plt.subplots(nrows=3, ncols=1, figsize=(10,10), sharex=True)
axs = axs.flatten()

axs[0].plot(data_raw[0][5000:], data_raw[1][5000:,:,0])
axs[1].plot(t_sub_2007b[2:], data_sub_2007b[2:, :])
axs[2].plot(ts_tvb[2:], data_bold_tvb[2:, :])

plt.show()
\end{lstlisting}

\end{document}